\documentclass[superscriptaddress,showpacs,amsmath,amssymb,aps,prd,twocolumn,floatfix,]{revtex4-1}

\usepackage{graphicx}
\usepackage{dcolumn}
\usepackage{bm}
\usepackage{amsmath}
\usepackage{amssymb}
\usepackage{subfigure}
\usepackage{lineno}
\usepackage{hyperref}
\usepackage{hhline}

\begin{document}

\title{A Deep Neural Network for Pixel-Level Electromagnetic Particle Identification in the MicroBooNE Liquid Argon Time Projection Chamber}

\newcommand{\Bern}{Universit{\"a}t Bern, Bern CH-3012, Switzerland}
\newcommand{\BNL}{Brookhaven National Laboratory (BNL), Upton, NY, 11973, USA}
\newcommand{\Cambridge}{University of Cambridge, Cambridge CB3 0HE, United Kingdom}
\newcommand{\Chicago}{University of Chicago, Chicago, IL, 60637, USA}
\newcommand{\Cincinnati}{University of Cincinnati, Cincinnati, OH, 45221, USA}
\newcommand{\CSU}{Colorado State University, Fort Collins, CO, 80523, USA}
\newcommand{\Columbia}{Columbia University, New York, NY, 10027, USA}
\newcommand{\FNAL}{Fermi National Accelerator Laboratory (FNAL), Batavia, IL 60510, USA}
\newcommand{\Harvard}{Harvard University, Cambridge, MA 02138, USA}
\newcommand{\IIT}{Illinois Institute of Technology (IIT), Chicago, IL 60616, USA}
\newcommand{\KSU}{Kansas State University (KSU), Manhattan, KS, 66506, USA}
\newcommand{\Lancaster}{Lancaster University, Lancaster LA1 4YW, United Kingdom}
\newcommand{\LANL}{Los Alamos National Laboratory (LANL), Los Alamos, NM, 87545, USA}
\newcommand{\Manchester}{The University of Manchester, Manchester M13 9PL, United Kingdom}
\newcommand{\MIT}{Massachusetts Institute of Technology (MIT), Cambridge, MA, 02139, USA}
\newcommand{\Michigan}{University of Michigan, Ann Arbor, MI, 48109, USA}
\newcommand{\NMSU}{New Mexico State University (NMSU), Las Cruces, NM, 88003, USA}
\newcommand{\Otterbein}{Otterbein University, Westerville, OH, 43081, USA}
\newcommand{\Oxford}{University of Oxford, Oxford OX1 3RH, United Kingdom}
\newcommand{\PNNL}{Pacific Northwest National Laboratory (PNNL), Richland, WA, 99352, USA}
\newcommand{\Pitt}{University of Pittsburgh, Pittsburgh, PA, 15260, USA}
\newcommand{\StMarys}{Saint Mary's University of Minnesota, Winona, MN, 55987, USA}
\newcommand{\SLAC}{SLAC National Accelerator Laboratory, Menlo Park, CA, 94025, USA}
\newcommand{\Syracuse}{Syracuse University, Syracuse, NY, 13244, USA}
\newcommand{\TelAviv}{Tel Aviv University, Tel Aviv, Israel, 69978}
\newcommand{\Tennessee}{University of Tennessee, Knoxville, TN, 37996, USA}
\newcommand{\UTA}{University of Texas, Arlington, TX, 76019, USA}
\newcommand{\Tubitak}{TUBITAK Space Technologies Research Institute, METU Campus, TR-06800, Ankara, Turkey}
\newcommand{\Tufts}{Tufts University, Medford, MA, 02155, USA}
\newcommand{\VTech}{Center for Neutrino Physics, Virginia Tech, Blacksburg, VA, 24061, USA}
\newcommand{\Yale}{Yale University, New Haven, CT, 06520, USA}

\affiliation{\Bern}
\affiliation{\BNL}
\affiliation{\Cambridge}
\affiliation{\Chicago}
\affiliation{\Cincinnati}
\affiliation{\CSU}
\affiliation{\Columbia}
\affiliation{\FNAL}
\affiliation{\Harvard}
\affiliation{\IIT}
\affiliation{\KSU}
\affiliation{\Lancaster}
\affiliation{\LANL}
\affiliation{\Manchester}
\affiliation{\MIT}
\affiliation{\Michigan}
\affiliation{\NMSU}
\affiliation{\Otterbein}
\affiliation{\Oxford}
\affiliation{\PNNL}
\affiliation{\Pitt}
\affiliation{\StMarys}
\affiliation{\SLAC}
\affiliation{\Syracuse}
\affiliation{\TelAviv}
\affiliation{\Tennessee}
\affiliation{\UTA}
\affiliation{\Tubitak}
\affiliation{\Tufts}
\affiliation{\VTech}
\affiliation{\Yale}

\author{C.~Adams} \affiliation{\Yale}
\author{M.~Alrashed} \affiliation{\KSU}
\author{R.~An} \affiliation{\IIT}
\author{J.~Anthony} \affiliation{\Cambridge}
\author{J.~Asaadi} \affiliation{\UTA}
\author{A.~Ashkenazi} \affiliation{\MIT}
\author{M.~Auger} \affiliation{\Bern}
\author{S.~Balasubramanian} \affiliation{\Yale}
\author{B.~Baller} \affiliation{\FNAL}
\author{C.~Barnes} \affiliation{\Michigan}
% ...

\collaboration{The MicroBooNE Collaboration\footnote{microboone\_info@fnal.gov}} \noaffiliation



\begin{abstract}
  We have developed a convolutional neural network (CNN) that can make a pixel-level prediction of objects in image data recorded by a liquid argon time projection chamber (LArTPC) for the first time. We describe the network design, training techniques, and software tools developed to train this network.
The goal of this work is to develop a complete deep neural network based data reconstruction chain for the MicroBooNE detector. We show the first demonstration of a network's validity on real LArTPC data using MicroBooNE collection plane images. The demonstration is performed for stopping muon and a $\nu_\mu$ charged current neutral pion data samples.
\end{abstract}

\maketitle

\section{Introduction}
LArTPCs are capable of producing high-resolution images of particle interactions.  This is one of the main reasons LArTPCs are the technology of choice for several current and future neutrino research programs including the Short Baseline Program~\cite{SBN} (SBN) and Deep Underground Neutrino Experiment~\cite{DUNE} (DUNE). The MicroBooNE experiment~\cite{UBDetector}, as a part of the SBN program, studies the $\nu_e$-like low energy event excess observed by the MiniBooNE collaboration~\cite{MiniBooNE} using a LArTPC detector and the on-axis Booster Neutrino Beamline~\cite{BNB} (BNB) at Fermilab as the $\nu_\mu$ source.

A LArTPC consists of a homogeneous volume of liquid argon bounded by a cathode plane and an anode plane. When a charged particle traverses the sensitive region, ionization electrons and scintillation light are produced along its trajectory. The scintillation light is detected within several nano-seconds by an array of photomultiplier tubes (PMTs), providing a timing measurement. 
Ionization electrons take a few milliseconds to drift toward the anode under the applied electric field and are detected by the anode plane, which is equipped with a set of wire planes and charge-sensitive readout circuits.
The spatial separation within the wires on the anode plane determines the spatial resolution of the recorded two-dimensional (2D) projection images. 
The speed of drift electrons along with their longitudinal diffusion, as well as the electronics response and the sampling frequency of the signal waveform determine the spatial resolution along the drift direction.

\begin{figure}[t]
 \centering
  \includegraphics[width=0.48\textwidth]{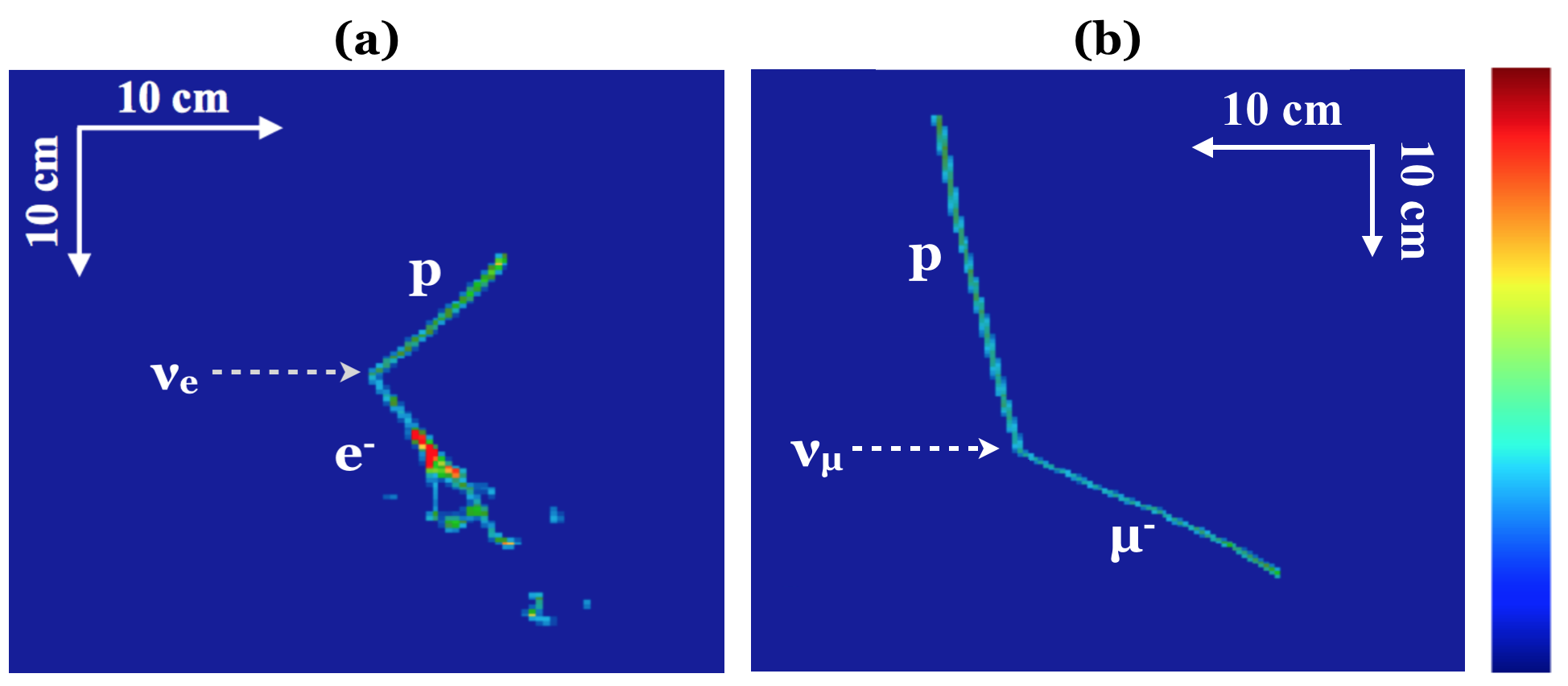}
 \caption{ Two examples of $\nu_e$ (a) and $\nu_\mu$ (b) events simulated in the MicroBooNE detector. The pixel color represents the amount of energy deposited per pixel in the 2D projection image. The pattern of ionization differs significantly between $\nu_e$ and $\nu_\mu$.}
  \label{fig:exampleimg}
\end{figure}

While the detector is able to capture many of the fine details that will be useful for physics analysis, parsing these details using automated reconstruction tools is still a technical challenge. One particular challenge is the discrimination between electromagnetic (EM) particles, namely $e^-$, $e^+$, and photons, and  other particle types. EM particles above the critical energy ($\approx$~33~MeV in argon) initiate an electromagnetic cascade of particles and develop a unique topology that consists of a collection of branching features. Identifying this topology is a simple form of particle identification, and is key information that can be used to discriminate between $\nu_e$ and $\nu_\mu$ interactions, as shown in Figure~\ref{fig:exampleimg}. For EM particles below the critical energy, other unique features are available.
One example is low energy electrons ($\delta$ rays) knocked off from argon atoms by a muon traversing the detector. These add multiple short {\it branches} to a muon trajectory. Separating them from a muon helps to identify the trunk of a muon trajectory. 
Another example is a Michel electron trajectory with low energy deposition per unit length ($dE/dx$) from the decay of a stopping muon. That trajectory typically exhibits a pattern of increasing $dE/dx$ due to an increase in ionization energy-loss as it comes to rest.
Identifying and cleanly separating energy depositions from Michel electrons in the muon decay allows one to reconstruct the Michel spectrum with high accuracy. This sample proves to be a valuable energy calibration source.
\begin{figure*}[t]
\centering
\includegraphics[width=0.9\textwidth]{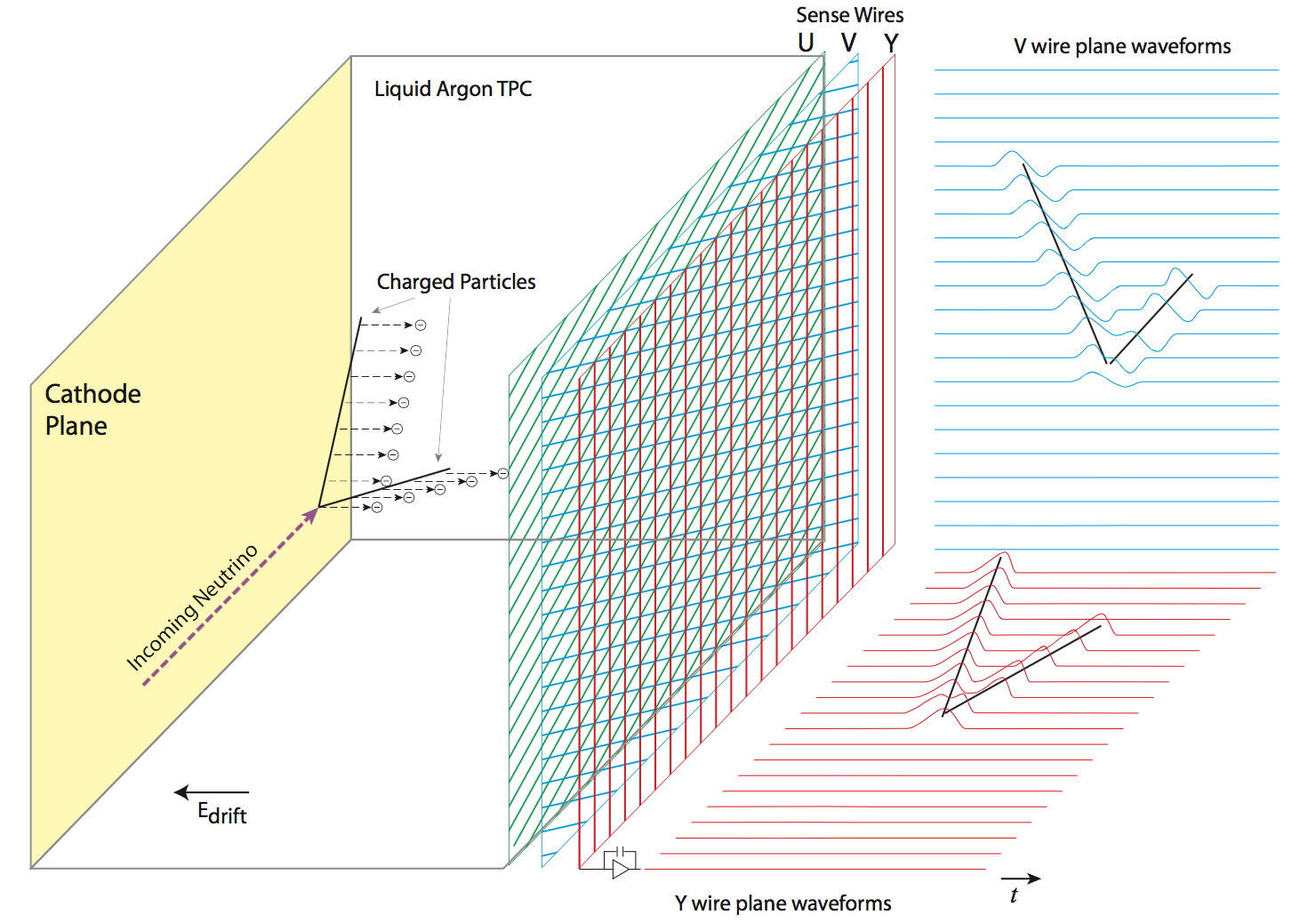}
\caption{ The MicroBooNE detector schematics showing three wire planes and example 2D projections of V and Y plane waveforms.
\label{fig:UBDetector}}
\end{figure*}

In this paper, we demonstrate that EM particles can be discriminated from other particles at the pixel-level in an image using a class of machine learning algorithms known as convolutional neural networks (CNNs).
We refer to EM particles as {\it shower} and the others as {\it track} particles in the rest of the paper. Our method uses noise-filtered, signal shape deconvolved waveforms~\cite{UBNoisePaper} from the charge-collection plane organized in 2D image format prior to a pixel clustering analysis. Having a prediction of each pixel as track or shower type prior to a pixel clustering reconstruction stage simplifies the downstream pattern-recognition algorithms and reduces the need for iterative reconstruction processes. We use a class of CNNs called {\it semantic segmentation networks} (SSNets) to classify each pixel of an image into a predefined set of semantics including a track, a shower, and a background pixel.

This work was performed as a step towards the development of a complete LArTPC event reconstruction and analysis chain using deep neural networks. It is an extension of an earlier study in which we demonstrated the use of image classification and object detection techniques with CNNs for LArTPC data analysis~\cite{UBCNNPaper}. The new contributions in this work include:
\begin{itemize}
\item First demonstration of a pixel-level object prediction for LArTPC event reconstruction using a deep neural network.
\item First demonstration of a deep neural network on real LArTPC detector data.
\item Software tools and algorithms capable of generating a pixel-level {\it label} (i.e., semantics) from {\sc{LArSoft}}~\cite{larsoft062601} data for supervised training and analysis of real detector data.
\end{itemize}
The software is open-source and accessible at Ref.~\cite{LArCV}.

The deep neural network described here is currently employed in the data reconstruction chain for the deep-learning based analysis of the MiniBooNE low-energy excess by MicroBooNE. The analysis goal is to locate a neutrino interaction vertex and identify the interaction type as being $\nu_e$ or $\nu_\mu$ for low energy neutrinos ($\approx$~200 to 600 MeV).  
Having track/shower separation information is crucial for vertex reconstruction and particle clustering algorithms. 

High rate environments, as in LArTPCs like the Short Baseline Near Detector (SBND) and the DUNE near detector, both to be built in the not-too-distant future at Fermilab, will benefit from sophisticated computer vision techniques. Such techniques, including pixel labeling as provided by the semantic segmentation approach described here, will be very useful to untangle neutrino-induced tracks and showers.

\begin{figure*}[t]
\centering
\includegraphics[width=0.85\textwidth]{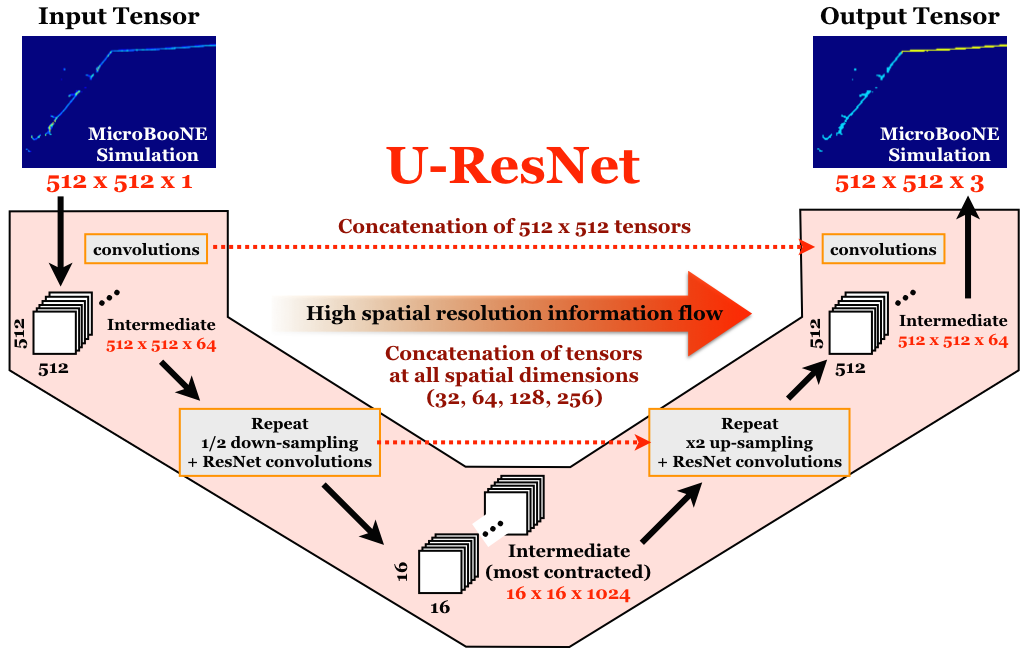}
\caption{U-ResNet architecture diagram. Black arrows describe the direction of tensor data flow. Red arrows indicate concatenation operations to combine the output of convolution layers from the encoding path to the decoding path. The final output has the same spatial dimension as the input with a depth of three, representing the background, track and shower probability of each pixel.
\label{fig:UResNet}}
\end{figure*}
\section{MicroBooNE Detector and Particle Images}


The MicroBooNE LArTPC contains 85 tonnes of liquid argon in the active region, which is defined by a rectangular shape with the dimensions 10.36~m in length, 2.32~m in height, and 2.56~m in width along the drift direction~\cite{UBDetector}, as shown in Figure~\ref{fig:UBDetector}.
The anode consists of three planes of parallel wires. The first and second planes contain 2,400 wires with orientations of +60 and -60 degrees from the vertical, respectively. 
Ionization electrons produce bipolar signals on the two {\it induction} planes as they pass through them.
The third is called the {\it collection} plane and consists of 3,456 vertical wires. Wires on the third plane are held at a positive potential and collect ionization electrons. Wires are separated by 3~mm pitch in all planes, and signal waveforms are digitized at a 2~MHz sampling rate and recorded for a duration of 4.8~ms in each event. 
Combined wire waveforms, aligned by the digitization time, form 2D projected images of a three-dimensional (3D) particle trajectory from a different projection angle. The digitization time runs along the vertical axis and the wires run along the horizontal axis in event displays shown in this paper (e.g. Figure~\ref{fig:exampleimg}).

In this paper we focus on the analysis of image data recorded by the collection plane, which has a size of 3,456 by 9,600 pixels. The spatial resolution of an image along the wire axis is 3~mm per pixel. For the analysis, every 6 samples of a digitized waveform are summed together, corresponding to an approximate spatial resolution along the time axis of 3.3~mm. The resulting image dimension is 3,456 by 1,600 pixels.



\section{U-ResNet: Track/Shower Pixel-level Separation Network}
In this study we use U-ResNet, a hybrid of the U-Net~\cite{UNet} and residual network~\cite{ResNet} (ResNet) design pattern. U-ResNet takes a single-channel 512 by 512 pixel image as input and outputs an image of the same spatial dimension with 3 channels per pixel encoding a probability from multinomial logistic regression, or {\it softmax}, for a pixel being a background, track, or shower type.
We use U-Net as the base SSNet architecture design because of its excellent performance in biomedical images~\cite{UNet} which resemble those from LArTPCs where information density is sparse.
We replace the convolution layers in the original U-Net with ResNet modules. ResNet is a generic CNN design pattern that was invented at the same time as U-Net and enables the training of deep CNNs. In our implementation, each ResNet module consists of two convolution layers of 3-by-3 kernel size, where each convolution layer is followed by a batch normalization operation~\cite{batchnorm} and a rectified-linear unit (ReLU) activation function. The schematic U-ResNet design is shown in Figure~\ref{fig:UResNet}.

The U-ResNet architecture can be interpreted in two separate sections. 
The first half of the network takes an input image, a data tensor with a dimension of (512,512,1), and repeatedly applies convolution and down-sampling operations. At the end of the first section, the data tensor has a dimension of (16,16,1024). The goal of this section of the network is to learn a non-linear, hierarchical representation of image features at different scales. 
Since feature information is encoded in a low spatial resolution tensor at the end of this section of the network, it is referred to as an {\it encoding path}.

The second half of the U-ResNet takes the output of an encoding path, a tensor with a dimension of (16,16,1024), and repeatedly applies an up-sampling and convolution operation. An up-sampling is performed by an operation called {\it convolution-transpose} which is an interpolation filter that expands the spatial dimension of a tensor by a factor of two. By using neurons to incorporate the interpolation operations, the U-ResNet architecture introduces a {\it learnable} interpolation filter optimized for assigning object classification to the pixels. This section of the network does essentially the opposite of the encoding path, hence it is called a {\it decoding path}.

An important and unique design pattern of U-Net is an additional path to allow the flow of information between the encoding path and the decoding path. This mitigates the loss of spatial resolution information in the encoding path where down-sampling is performed. The idea and method employed in U-Net is simple and effective; in the decoding path where high spatial resolution needs to be recovered, we concatenate the data tensors from the encoding path where the spatial dimension matches.  In the encoding path, data tensors hold the best possible spatial resolution at each spatial resolution prior to the down-sampling operations. Thus a simple concatenation allows information about spatial resolutions to flow into the decoding path, allowing U-Net to perform a high-precision image segmentation.

\section{Training U-ResNet}
We train U-ResNet via a supervised learning method that uses simulated particle interaction images.  In this section we describe techniques employed for the training and optimization methods.

\subsection{Transfer Learning}
We exploit a {\it transfer learning} technique by first training the first-half of U-ResNet for an image classification task using the identical data set from our previous publication~\cite{UBCNNPaper}. This data set contains single particle images which could be  $e^-$, photon, muon, $\pi^-$ or proton. The network's weights trained to discriminate between different particle images provide a natural initial state to perform a pixel-level track/shower separation. 
When we subsequently train the whole U-ResNet with pre-trained weights, we let all network parameters be trained and fine-tuned.

\subsection{Class/Pixel-wise Loss Weighting}
Training of the U-ResNet is a process of minimizing the loss, a measure of an error made by the network, over many iterations. The loss is computed by summing over a pixel-wise multinomial logistic loss in each image, and then averaging over all images in a batch of images. This definition of loss presents a challenge to training U-ResNet for LArTPC images where the fraction of pixels that are background is 99\% or more, hence dominating the total loss. In order to mitigate this challenge, the authors of the original U-Net paper introduced a class-wise loss (CL) weighting factor~\cite{UNet} which is a reciprocal for the number of pixels that belong to each class in an image. 

\begin{figure}[t]
	\vspace{0.1in}
    \centering
    \includegraphics[width=0.48\textwidth]{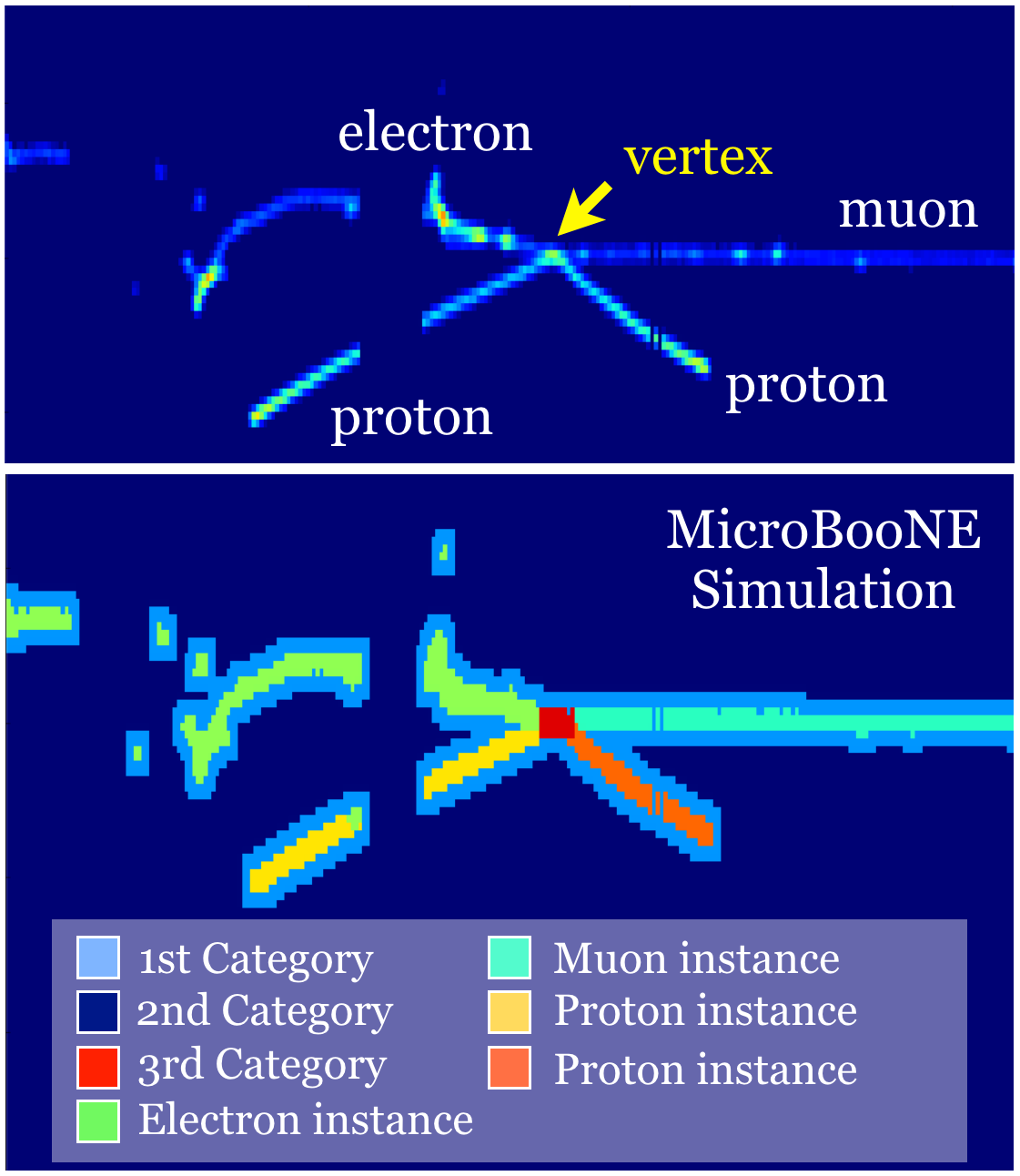} \\
\caption{Top: an example image from the training set in which two protons, one electron, and one muon are produced. 
The gaps along the trajectory of an electron and proton on the left are due to unresponsive wires~\cite{UBNoisePaper} in the detector. Bottom: the event from the top image that shows PL weighting categories indicated in different colors.  }
\label{fig:LossWeight}
\end{figure}

In this study, we introduce a pixel-wise loss (PL) weighting factor that is multiplied by a pixel's loss contribution to the total loss of an image. PL weighting enables the network's training to focus on challenging parts of an image by up-weighting a pixel loss in the corresponding regions. 
For the calculation of PL weighting factors, we define four categories of pixels with the last category separated into particle type instances. 
The first category contains background (i.e., zero) pixels that surround non-background (i.e., non-zero) pixels within 2 pixels. 
The second category is the rest of background pixels in the image that do not belong to the first category. 
The third category represents non-zero pixels within 4 pixels of the generated event vertex. 
Finally, the fourth category is defined for each particle instance and includes non-zero pixels that belong to a particle. 
Therefore the total number of categories may vary from one event to another.
A PL weighting factor is computed per category. It is the reciprocal of the number of pixels belonging to each category. A category with fewer pixel counts represents a rare feature in an image data, and is assigned a higher weighting factor.  Figure~\ref{fig:LossWeight} shows how pixels are grouped into the four categories.

\subsection{Optimization}
\begin{figure}[t]
\center
\includegraphics[width=0.4\textwidth]{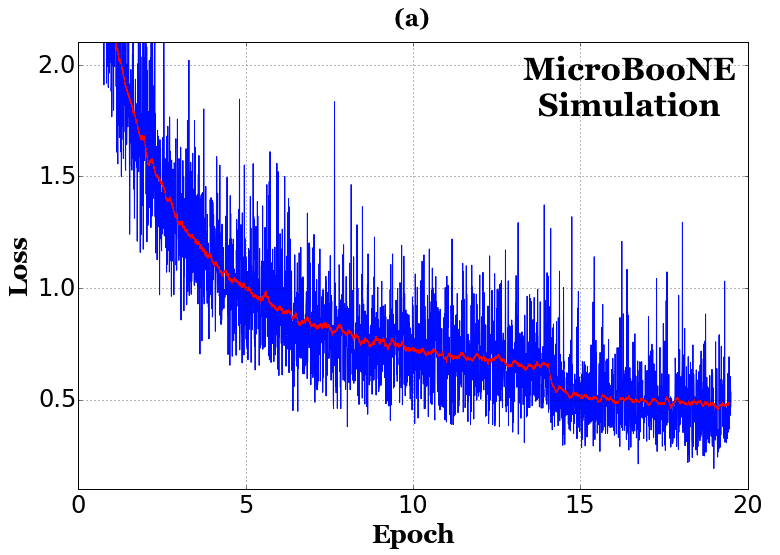}\\
\vspace{0.1in}
\includegraphics[width=0.4\textwidth]{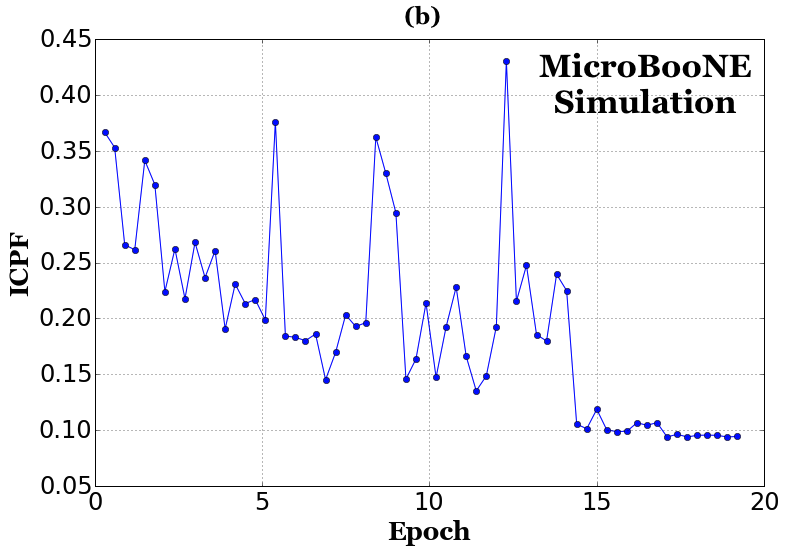}
\vspace{0.1in}
\caption{(a) The training loss value as a function of training time using the validation sample.  The red line shows the average at a given Epoch computed using 200 the neighboring Epoch points. (b) Incorrectly Classified Pixel Fraction (ICPF) for the same sample. The sudden drop in both figures at Epoch 14 is due to lowering of the learning rate by a factor of 10.}
\label{fig:EpochPlot}
\end{figure}

We use the {\sc{RMSProp}}~\cite{rmsprop} with an initial learning rate of 0.0003 to optimize the U-ResNet.
The weights are updated after processing every batch of 60 images. 
The training process is monitored using the Incorrectly Classified Pixel Fraction (ICPF) metric. Figure~\ref{fig:EpochPlot} shows the loss and the average values of ICPF computed over validation samples as a function of epoch during the training. Epoch is a measure of time, and one epoch corresponds to the time it takes to consume the same number of images as the whole training sample. 
The learning rate is lowered by an order of magnitude at epoch 14 as shown in Figure~\ref{fig:EpochPlot}. 
We determine the best performing network parameters based on the lowest ICPF value on the {\it validation set} which is generated independently from the data set used for training under the same simulation configuration. The performance of the network is then quantified using the {\it test set}, which is yet another independent sample generated with the same simulation configuration. Using the trained network, we find an average ICPF value of 1.9\% using all events in the test set.

The U-ResNet and this training scheme is implemented using {\sc{caffe}}~\cite{caffe}, customized to employ the PL weighting scheme~\cite{LArCV}.  We trained our network using NVIDIA TitanX~\cite{titanx} GPUs with 12 GB memory.

\section{Data Samples for Training and Validation}

\subsection{Training Sample Preparation}
We prepare training samples using a custom event generator called {\sc{MultiPartVertex}} (MPV), available in the MicroBooNE software repository, {\sc{uboonecode}}~\cite{uboonecode06260113}. MPV can be configured to randomly generate a single 3D point in a detector with the emission of multiple charged particles. Any random process employed by MPV is a uniform distribution within the specified range in the configuration. 
The multiplicity and type of particles to be generated are configurable parameters as outlined below. Restrictions and ranges for the generation are presented in the following two paragraphs.

For 80\% of the sample, the MPV is configured to generate events with a random total particle multiplicity between one and four. One of the generated particles must be a light lepton ($e^-$ or $\mu^-$) with kinetic energy ranging from 50 to 1000 MeV. The direction of each particle is chosen from an isotropic distribution. For the other generated particles, the MPV is configured to randomly assign their types to a photon, charged pion, proton, or another lepton ($e^-$ or $\mu^-$). 
We also set the maximum multiplicity for leptons and protons to be three and photons and charged pions to be two.
There is no strong motivation for this configuration. 
In fact we demonstrate later in this paper that the network works well on neutrino candidate events with a shower particle from real detector data with multiplicity five.

The remaining 20\% of images are generated with a different configuration. The total multiplicity is set randomly between one and four particles but there is no restriction to include at least one light lepton. Instead, particle types are set randomly between showers ($e^-$ and photon) and tracks ($\mu^-$, charged pion, and proton). For each particle type, the maximum multiplicity is set to two. 
The ranges for the randomly assigned momentum are specified as 30 to 100~MeV/c for $e^-$ and photon, 85 to 175~MeV/c for $\mu^-$, 95 to 195~MeV/c for charged pion, and 300 to 450~MeV/c for proton. The distribution of particle directions is isotropic. This 20\% fraction is chosen to have a particular focus on the low energy region where classification of particle types becomes difficult. 
The motivation for this is to enhance the network’s performance in this energy region.
We generate 140,000 images, randomly selecting 100,000 for training, 20,000 for validation, and 20,000 images for testing the network's performance, with no image in more than one set.

After the particle generation stage, the training sample is run through the detector simulation and waveform processing scheme of the experiment.
The procedure is similar to that of our previous study~\cite{UBCNNPaper} but with an updated version of {\sc{LArSoft}}~\cite{larsoft062601} and {\sc{uboonecode}}~\cite{uboonecode06260113}. The latter software contains important updates including a data-driven detector noise model~\cite{UBNoisePaper}, noise filtering algorithms, and data-driven TPC charge signal deconvolution kernels~\cite{UBSignalPaper1,UBSignalPaper2}. 
These improvements aim to reduce the potential discrepancies between data and simulation samples.
Hence, these improvements are important for the U-ResNet trained purely on simulation to work effectively on real detector data. 
Further suppression of the discrepancy between data and simulation for noise with low amplitude is accomplished by setting the pixel value to zero for pixels with amplitude below 10.
Finally, we crop the 512 by 512 pixel image from the whole collection plane image which has the original size of 3,456 by 1,600 pixels. 
 The cropping algorithm (CRA) defines an axis-aligned 3D rectangular volume within the detector of a configurable size that contains a set of 3D points, called constraint points.
The location of the 3D box is set by the algorithm under two conditions. First, the defined box must contain all given 3D constraint points. Second, while satisfying the first condition, a range is defined that maximizes the number of non-zero 2D pixels included in the projection of the rectangular box in the collection plane.  By satisfying these two conditions, the box location is allowed to float freely. We use the 3D interaction vertex as the constraint point in this study. 
The resulting 512 by 512 pixel images contain the interaction point location for each event and the maximum number of non-zero pixels in the projection.

\subsection{Benchmark Simulation Samples}
\label{sec:simulation_sample}
Separately from the testing set, we generate five additional simulation samples to benchmark the performance of U-ResNet. These simulation samples include two types of neutrino interactions simulated using the {\sc{GENIE}}~\cite{genie} neutrino event generator within {\sc{LArSoft}}~\cite{larsoft062601} and the MPV generator events generated under three different generator configurations. The image preparation steps are identical to those of the training samples except for the event generation step which is unique to the generator type and configuration. 
This brings us to a total of six simulated samples, consisting of 120,000 events, that we can analyze with the trained network.

The neutrino samples consist of 20,000 $\nu_\mu$ and 20,000 $\nu_e$ events, generated with the Booster Neutrino Beam~\cite{BNB} (BNB) beam flux information. 
Each MPV samples includes 20,000 images of events. One MPV sample is configured to generate one proton and one electron only (1e1p). Particles are simulated with a uniform energy distribution and isotropic momentum direction distribution. The kinetic energy range is set to be 50 to 500~MeV for $e^-$ and 50 to 300~MeV for protons. 
In addition, there are two more MPV samples generated: low energy 1e1p (1e1p-LE) and low energy 1$\mu$1p (1$\mu$1p-LE) where the latter is similar to 1e1p except a $\mu^-$ is generated in place of an $e^-$. These samples are generated in the low energy (LE) range. For 1e1p-LE, the $e^-$ has momentum distributed from 30 to 100~MeV/c.  For 1$\mu$1p-LE, the momentum of $\mu^-$ is distributed  from 85 to 175~MeV/c. For both samples, the momentum of the proton is distributed from 300 to 450~MeV/c. 

\subsection{Benchmark Data Samples}
In order to validate the network's performance on real data, we prepared two data samples for which we have a good understanding from the traditional reconstruction approaches available in LArSoft~\cite{UBPandoraPaper}. 

The first is a sample of Michel electron events~\cite{UBMichelPaper}, also used in our first physics result publication. This sample primarily consists of one track (stopping muon) and one shower (decay electron) and is identified using a reconstruction algorithm developed by the collaboration. The Michel electron images are simple and therefore useful to study how the network response depends on a limited amount of image features. 

The second data set is a sample of charged-current $\nu_\mu$ candidate interactions with one or more photons produced, primarily via $\pi^0$ decay at the interaction vertex. 
The CC$\pi^0$ sample gives a different perspective than Michel electrons because it primarily consists of higher energy showers and tracks that make the image more feature-rich and complicated. Validation of the network performance on both data sets is crucial.

For the Michel electron sample we use a random subset of events identified as Michel electron events in Ref~\cite{UBMichelPaper}. We processed 100 data and 100 simulation events through the same waveform processing procedures applied to generate our training sample. Then we use the reconstructed decay position of the Michel electron as a constraint to crop with the CRA, which produce 512 by 512 pixel images containing both a stopping muon and a decay electron. Next we use the {\sc{LArCV}}~\cite{LArCV} toolkit to produce a pixel-level categorization of track and shower pixels through hand-scanning of images by physicists. In this process, we ignored pixels that are related neither to a stopping muon nor a Michel electron. The ignored pixels are typically due to other cosmic ray muons or secondaries produced by them. 
This allows us to reduce the number of pixels to be labeled.
The disagreement rate between the physicist-labels and the U-ResNet's classification is then compared between the real data and simulated data to quantify how the network performance differs between data and simulation.

For the CC$\pi^0$ events, samples of 100 data images are identified primarily by an automated reconstruction~\cite{UBPandoraPaper}. 
Event selection algorithms look for a $\nu_\mu$ CC interaction candidate vertex, namely a muon track with EM-showers from $\pi^0$ decay near to that vertex.  Such a muon track must be either contained or associated with a proton track to reject cosmogenic backgrounds. Selected events must pass through a subsequent hand-scanning process by physicists to ensure a high purity.
The reconstruction algorithm in the reconstruction chain provides an estimate of the interaction vertex position. This reconstructed vertex is used as a constraint for CRA to produce 512 by 512 pixel images. For the comparison study, we simulate events with BNB $\nu_\mu$ interactions and cosmic rays. A total of 100 CC$\pi^0$ events are selected based on simulation information. The neutrino interaction vertex location from simulation information is used as a constraint for CRA to produce the same size images for simulated events. Data and simulation events are processed by an identical waveform process chain used to prepare the training sample. Finally, the pixel-level physicist labels are generated for the CC$\pi^0$ sample. The same condition is applied and the physicists labeled only pixels that are considered to be related to a neutrino interaction, ignoring pixels with cosmic ray induced energy depositions. 

\section{Network Performance on Simulation Samples}
\begin{figure}[t]
  	\includegraphics[width=0.48\textwidth]{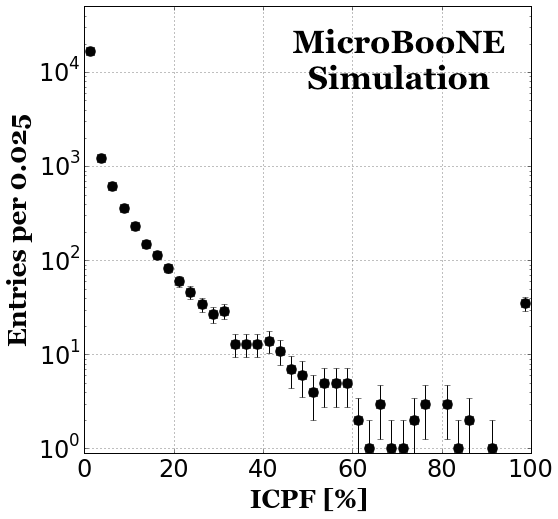}
\caption{The binned ICPF distribution over all images in the test set.}
\label{fig:ICPF}
\end{figure}

We benchmark the performance on test simulation samples using four metrics.
\begin{itemize}
\item ICPF mean: the average value of incorrectly classified pixel fraction per image computed over all events in a sample. The ICPF metric is a measure that takes into account false positives and the fraction of labels for the track and shower categories.
\item ICPF $90\%$ quantile: the ICPF value below which 90\% of events in a sample are present.
\item Shower error rate: the average value of the shower pixel error rate, defined as the fraction of incorrectly labeled shower pixels as track pixels in each image, averaged over all images in a sample. 
\item Track error rate: the average value of the track pixel error rate, defined as the fraction of incorrectly labeled track pixels as shower pixels in each image, averaged over all images in a sample. 
\end{itemize}
For all samples, the ICPF distributions are very similar. We show one example for the test sample in Figure~\ref{fig:ICPF}. 
In general, most images have very low ICPF values, well below $10\%$ for all test samples.

The results can be found in Table~\ref{tab:TestAccuracy}. The network is generalized to perform well on simulated neutrino events to a level that allows us to apply the technique as a part of the reconstruction chain. We do not train the network on our signal prediction -  neutrino events simulated by {\sc{GENIE}} - because this may introduce a model bias. The benchmark results also demonstrate that the U-ResNet can classify pixels from the low energy two particle topologies of 1e1p and 1$\mu$1p into track/shower at the  ICPF mean value of $3.9\%$ and $2.3\%$, respectively. These are the two simplest topologies of neutrino interactions, and it is important for U-ResNet to perform well so that it can be used to distinguish the two neutrino flavors. 
In the 1$\mu$1p-LE sample, despite the fact that no showers are produced in the primary neutrino interaction, challenges for the network arise from similarities between muons and electrons at very low energies and from secondary interactions like Michel electrons from muon decays.
\begin{table}[t]
\begin{center}
\caption{Values of the network performance metrics including the average of ICPF mean value, 90\% quantile, the average of incorrectly classified pixel fraction for shower pixels, and the same for track pixels. The test samples described in Section~\ref{sec:simulation_sample}. Values are shown in percentages.}
\vspace{0.1in}
\begin{tabular}{lcccc}
\hline
\hline
 & {\bf ICPF } & {\bf ICPF} & & \\
{\bf Sample}& {\bf mean} & {\bf 90\%} & {\bf Shower} & {\bf Track} \\
\hline
Test        & 1.9       & 4.6       & 4.1    & 2.6  \\
$\nu_e$     & 6.0       & 13.8      & 7.6    & 13.8 \\
$\nu_\mu$   & 3.9       & 4.5       & 14.2   & 4.3  \\
1e1p        & 2.2       & 5.7       & 2.8    & 4.0  \\
1$\mu$1p-LE & 2.3       & 2.2       & 6.2    & 2.4  \\
1e1p-LE     & 3.9       & 11.5      & 3.8    & 8.0  \\
\hline
\hline
\end{tabular}
\end{center}
\label{tab:TestAccuracy}
\end{table}

\begin{figure}[t]
  	\includegraphics[width=0.48\textwidth]{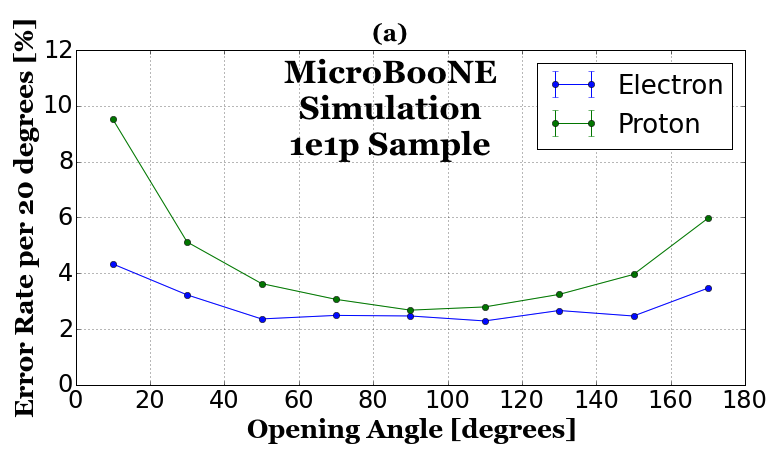}\\
    \vspace{0.1in}
    \includegraphics[width=0.48\textwidth]{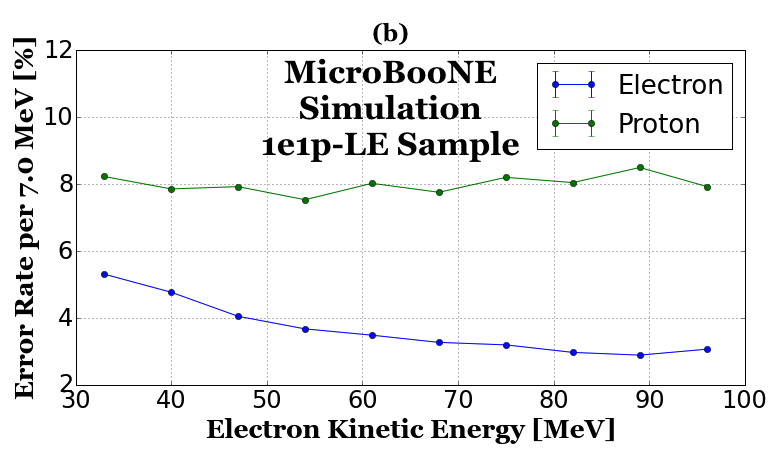}\\
    \vspace{0.1in}
   	\includegraphics[width=0.48\textwidth]{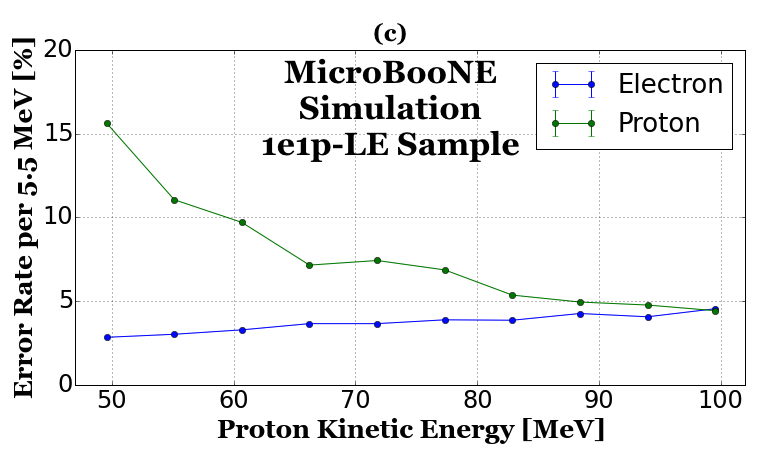}
\caption{ The ICPF error rates for U-ResNet labeling track (proton) and shower (electron) pixels for the benchmark test set are plotted against initial kinematic variables from simulation information. (a) The opening angle between two particles from the 1e1p sample. (b) The electron kinetic energy from the 1e1p-LE sample. (c) The proton kinetic energy from the 1e1p-LE sample.}
\label{fig:ErrorRate_1e1p_TestSample2}
\end{figure}

Figure~\ref{fig:ErrorRate_1e1p_TestSample2} shows the binned ICPF for the 1e1p and 1e1p-LE samples as a function of kinematic variables. Figure~\ref{fig:ErrorRate_1e1p_TestSample2}(a) shows the correlation with the opening angle between the two particles in 1e1p sample. 
We expect the ICPF value to increase when the two particles are co-linear and the 2D projections overlap, making it hard to distinguish the two tracks. When they are back-to-back, the difficulty to distinguish them arises from the fact that two trajectories may appear as the trajectory of one particle. 
Although it is outside of the scope of this paper, some of these difficulties could be mitigated if multiple 2D projection information is incorporated. 

Figure~\ref{fig:ErrorRate_1e1p_TestSample2}(b) and (c) show the dependence of the performance on the kinetic energy of a particle from 1e1p-LE sample. We observe that the network performs worse at lower energies. The ICPF value reaches near 15\% at 50~MeV proton kinetic energy. A proton at this energy can only travel a few centimeters in LAr, which translates into 10 pixels or fewer in the collection plane image. Such a small amount of information makes the network’s task difficult. A similar trend of decreasing performance can be also seen for electron kinetic energy, although the magnitude is much smaller. 
 The critical energy above which electrons primarily produce bremsstrahlung in LAr is about 33 MeV.
In the low energy region near or below the critical energy, electrons may not show a geometrical feature of showers characterized by a cascade of radiation. Thus, the network may struggle identifying them as showers. 

Overall, these kinematic distributions show the trend we expect, and set milestones to be achieved by future work on deep neural network development for LArTPC data reconstruction. A few randomly chosen example outputs of the networks are shown in the Appendix from the $\nu_e$ and $\nu_\mu$ benchmark set.

\section{Network Performance with Detector Data and Comparison to Simulation}
In this section we report the validation of U-ResNet on real detector data, in particular Michel electron and CC$\pi^0$ neutrino candidate events. Both data and simulation samples are processed by a physicist and contain pixel-level prediction labels. We report the comparison of the network's disagreement with physicist-applied labels. The details of data preparation steps are described in the previous sections.

\subsection{Data/Simulation Comparison Using The Michel Electron Sample}

\begin{figure}[t]
	\centering
    \includegraphics[width=0.45\textwidth]{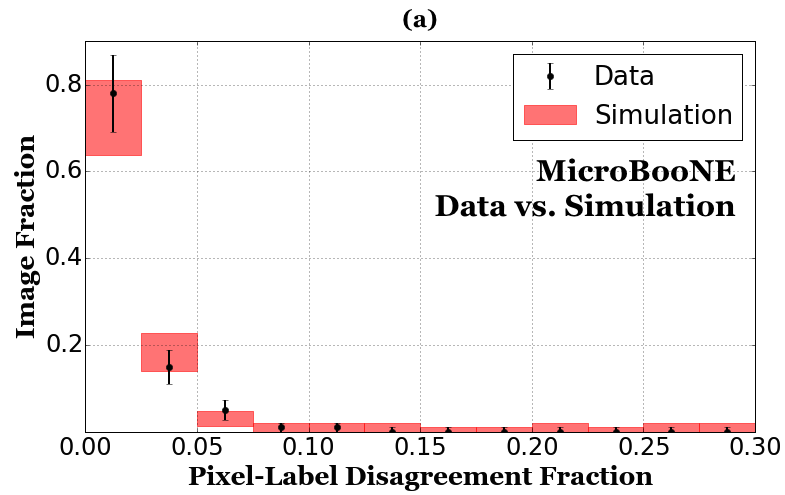}\\
    \vspace{0.1in}
	\includegraphics[width=0.45\textwidth]{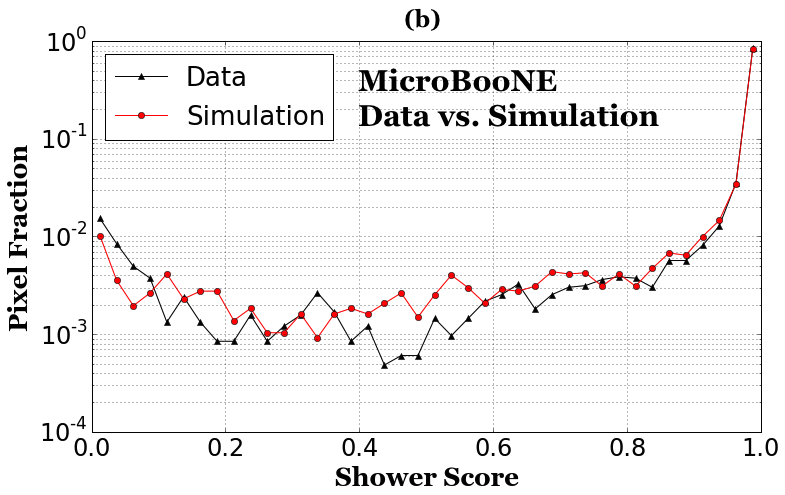}\\
    \vspace{0.1in}
    \includegraphics[width=0.45\textwidth]{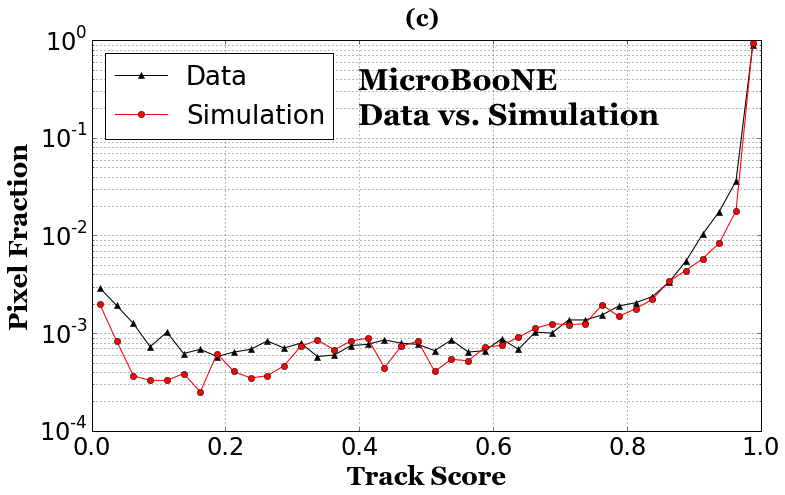}
	\caption{(a) Binned distribution of ICPF for track and shower pixels where the pixel-level labels are produced by a physicist. The data and simulation distributions are area-normalized and represent 100 Michel electron events. There is no event outside the shown range on the horizontal axis. (b) The normalized, binned softmax probability distributions for shower pixels on data and simulation. (c) The same as (b) for track pixels.}
	\label{fig:DataMC_Score_Michel}
\end{figure}

Table~\ref{tab:TestAccuracyMichel} summarizes the analysis results for the Michel electron sample. The disagreement rate between a physicist analyzer and the network prediction is below the 3\% level on average for both data and simulation. 
Figure~\ref{fig:DataMC_Score_Michel}(a) shows the distribution of the pixel fractions where U-ResNet and physicists disagreed on track/shower categorization over 100 events. 
The calculated physicist/network labeling for data and simulation agrees within statistical uncertainty. Figure~\ref{fig:DataMC_Score_Michel} also shows binned distributions of pixel scores for data and simulation. The track or shower label for each pixel is assigned by a physicist analyzer, and is not expected to be perfect. The score distributions show a similar trend between data and simulation. The error bars are not drawn in the score distributions since it is not trivial to derive an error for a pixel-wise score where we expect strong inter-pixel correlations. Finally, Table~\ref{tab:TestAccuracyMichel} shows that the network has a smaller ICPF when using labels generated from simulation information. 
The Appendix shows four randomly selected examples of Michel electron events from real data.
\begin{table}[t]
\begin{center}
\caption{Values of the network performance metrics for the Michel electron sample. The top row indicates the type of sample used (simulation or data), the second shows the source of a label used for analysis, and the third shows the source of a pixel prediction. The forth and the fifth rows indicate the ICPF mean value over all samples and 90\% quantile, respectively. The bottom two rows show the mean of ICPF for track and shower pixels, separately.  Values are given as percentages.}
\vspace{0.1in}
\begin{tabular}{l||c|c|c|c}
\hline
\hline
Sample    & Data     & Simulation & Simulation & Simulation \\
\hline
Label     & Physicist    & Physicist      & Simulation & Simulation \\ 
\hline
Prediction& U-Resnet & U-ResNet   & U-ResNet   & Physicist \\
\hline
ICPF mean & 1.8      & 2.6        & 2.5        & 2.3 \\
\hline
ICPF 90\% & 3.3      & 4.4        & 4.5        & 3.1 \\
\hline
Shower    & 6.2      & 5.7        & 4.0        & 3.9 \\
\hline
Track     & 1.1      & 1.9        & 1.6        & 1.3 \\
\hline
\hline
\end{tabular}
\label{tab:TestAccuracyMichel}
\end{center}
\end{table}

\begin{figure}[t]
\centering
\includegraphics[width=0.45\textwidth]{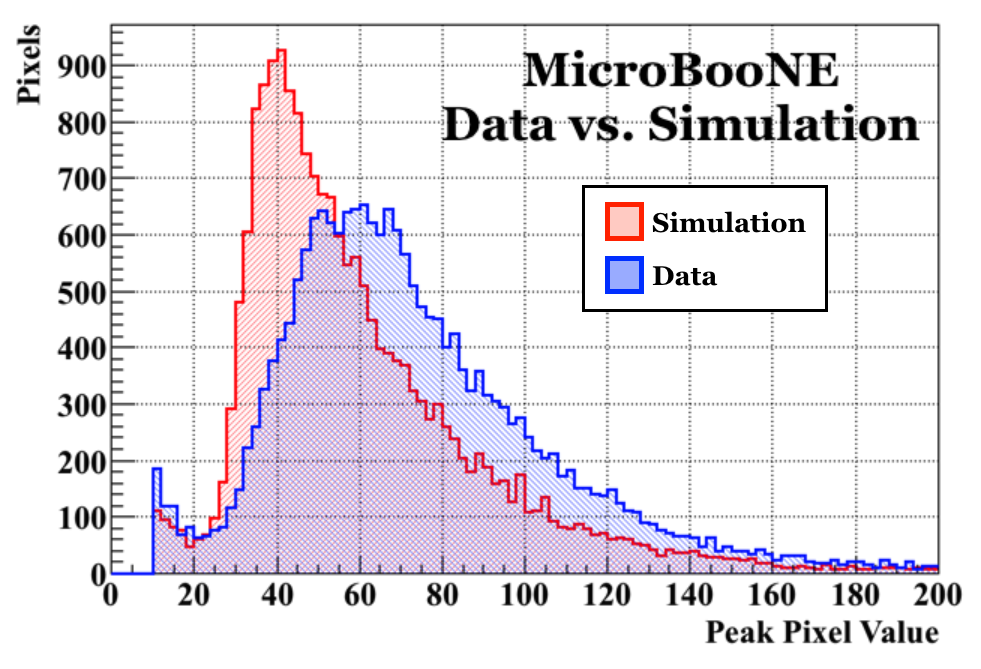}
\caption{Peak pixel value distribution for Michel electron images for data and simulation using the 3-pixel differentiation algorithm described in the text. The vertical axis shows the pixel counts while the horizontal axis shows the peak pixel values. 
\label{fig:MichelPeakADC}}
\end{figure}
\begin{figure}[t]
	\centering
	\includegraphics[width=0.45\textwidth]{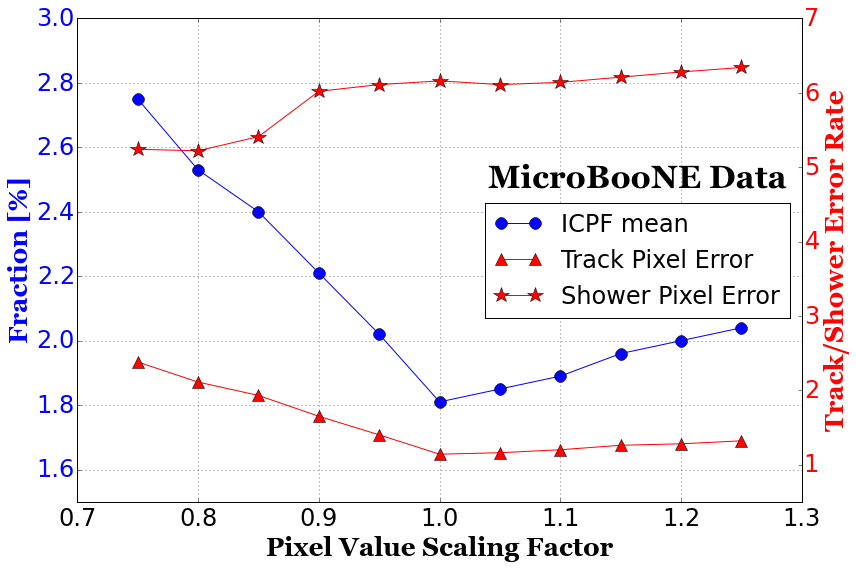}
	\caption{The average ICPF in percent for the Michel electron data versus pixel scaling factor is shown in blue.  The pixel-fraction disagreement between physicist and network categorizations  is shown in percent in red for track and shower pixels separately.}
	\label{fig:MichelScalingResponse}
\end{figure}

Further, we inspected the robustness of the network against scaling pixel values. Since there is no calibration applied at the stage of processed waveforms, we expect a difference in the signal strength between data and simulation. We run a simple differentiation algorithm to compare the signal strength between Michel data and simulation images. The algorithm inspects every pixel in an image. 
The algorithm finds peak pixels by comparing a given pixel with the one before and after it along the time axis to determine the one with a higher pixel value than its neighbors.
This is the simplest form of a signal peak amplitude finder algorithm. The distributions for data and simulation are shown in Figure~\ref{fig:MichelPeakADC}, which shows a shift between the data and simulation peak positions by about 20\% to 30\%. For this study we scale the pixel values of data images by a constant factor and compare the performance of the network with different scaling factors. Figure~\ref{fig:MichelScalingResponse} shows the results of this study.
Although we observe that the ICPF becomes worse when we apply the scaling factor, the change is within 1\% absolute when we scale pixel values by 25\%, which is at the level of current disagreement rate between simulation and uncalibrated detector response.


\subsection{Qualitative Analysis of Inter-Pixel Correlations Using the Michel Electron Sample}
\begin{figure}[t]
	\centering
	\includegraphics[width=0.4\textwidth]{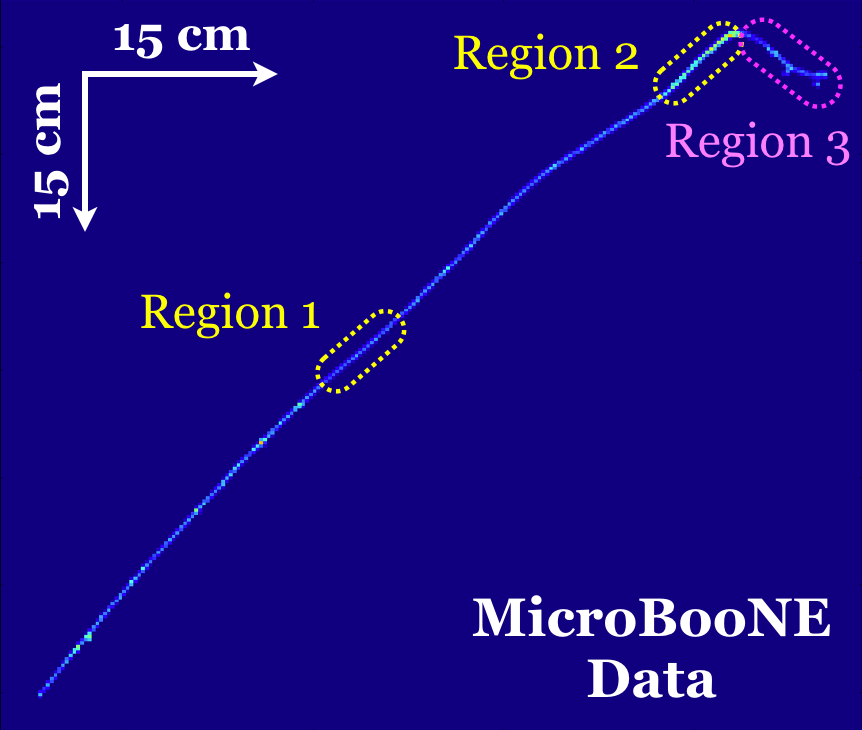}
    \caption{Three regions for an analysis of the inter-pixel correlation for a Michel in data. Region~1 contains the minimum ionizing muon trajectory. Region~2 focuses on the end of the muon trajectory where $dE/dx$ increases. Region~3 contains a decay Michel electron.}
	\label{fig:OneEventExample_Michel_01}
\end{figure}

We take a qualitative look at the correlation of pixel scores using the Michel electron image from one of the real data examples shown in Figure~\ref{fig:OneEventExample_Michel_01}. In the following sections, we focus on three regions shown in the figure. 
These correspond to 1) a minimum-ionizing muon track, 2) a portion of the track with high dE/dx near the muon stopping point, and 3) the low energy Michel electron shower, respectively.

\subsubsection{Minimum Ionizing Muon Track}
\begin{figure*}[t]
\centering
   	\includegraphics[width=1.0\textwidth]{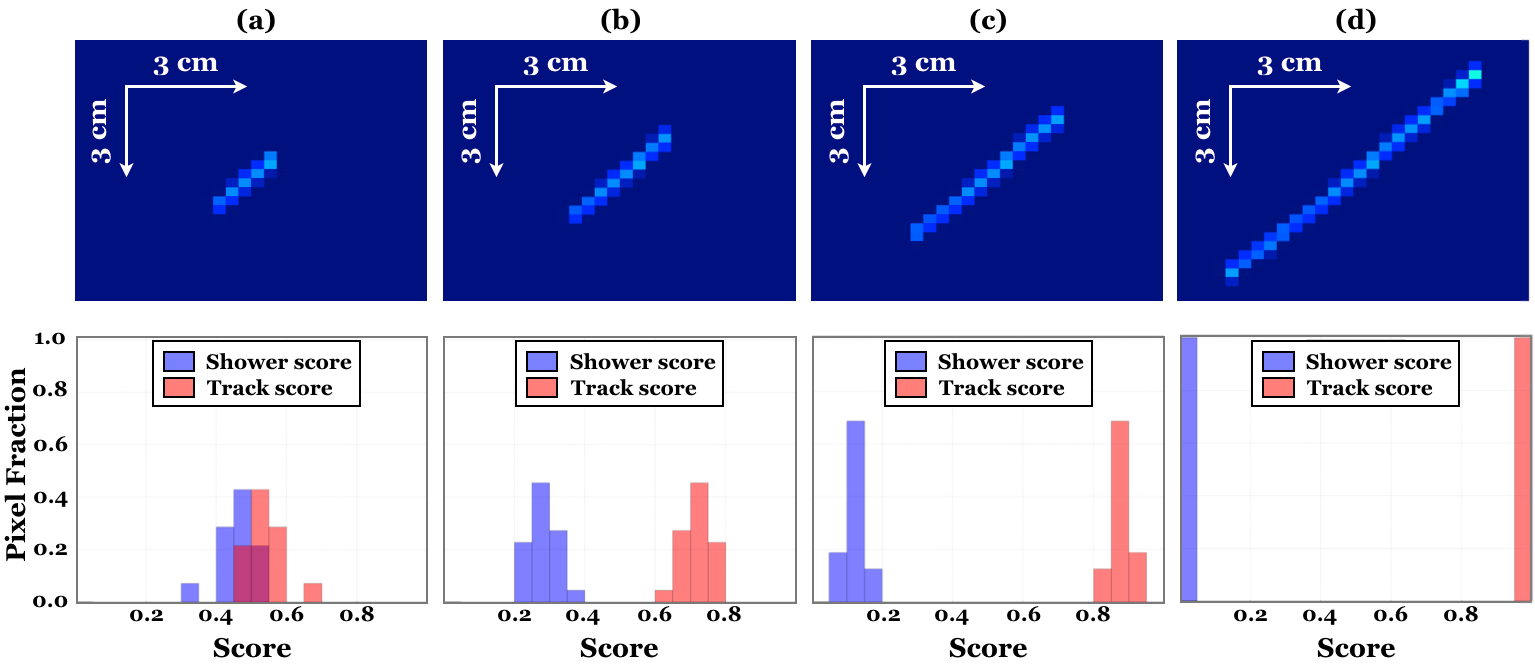}
\caption{Minimum ionizing muon track in Region~1 of Figure~\ref{fig:OneEventExample_Michel_01} where all pixels in the image are masked except for a small portion of a muon track shown in the images above. The lower row shows normalized track vs. shower score distributions for all non-zero pixels in the image. }
\label{fig:Region1_Michel}
\end{figure*}
One possible property used to distinguish a minimum ionizing muon from a low energy electron is the topology of its trajectory. This is often a long straight line, as compared to a more ``jagged'' electron trajectory due to higher multiple coulomb scattering. We choose subsets of Region~1 shown in Figure~\ref{fig:OneEventExample_Michel_01} to test this hypothesis by masking all remaining pixels in the image to zero. 
Figure~\ref{fig:Region1_Michel} shows the masked images and the corresponding track vs. shower score distribution of non-zero (i.e., unmasked) pixels by running the U-ResNet on each image. 
In Figure~\ref{fig:Region1_Michel} (a) to (d), we show a series of images with increasing number of unmasked pixels to determine how the score distribution changes. 
When we provide only a 5-pixel long minimum ionizing track, separation is weak. The separation improves as we include more neighboring pixels, which makes the straight-track shape longer and longer. We conclude that this confirms our hypothesis that the network is focusing on the length of a straight minimum ionizing particle's trajectory.


\begin{figure*}[t]
\centering
   	\includegraphics[width=1.0\textwidth]{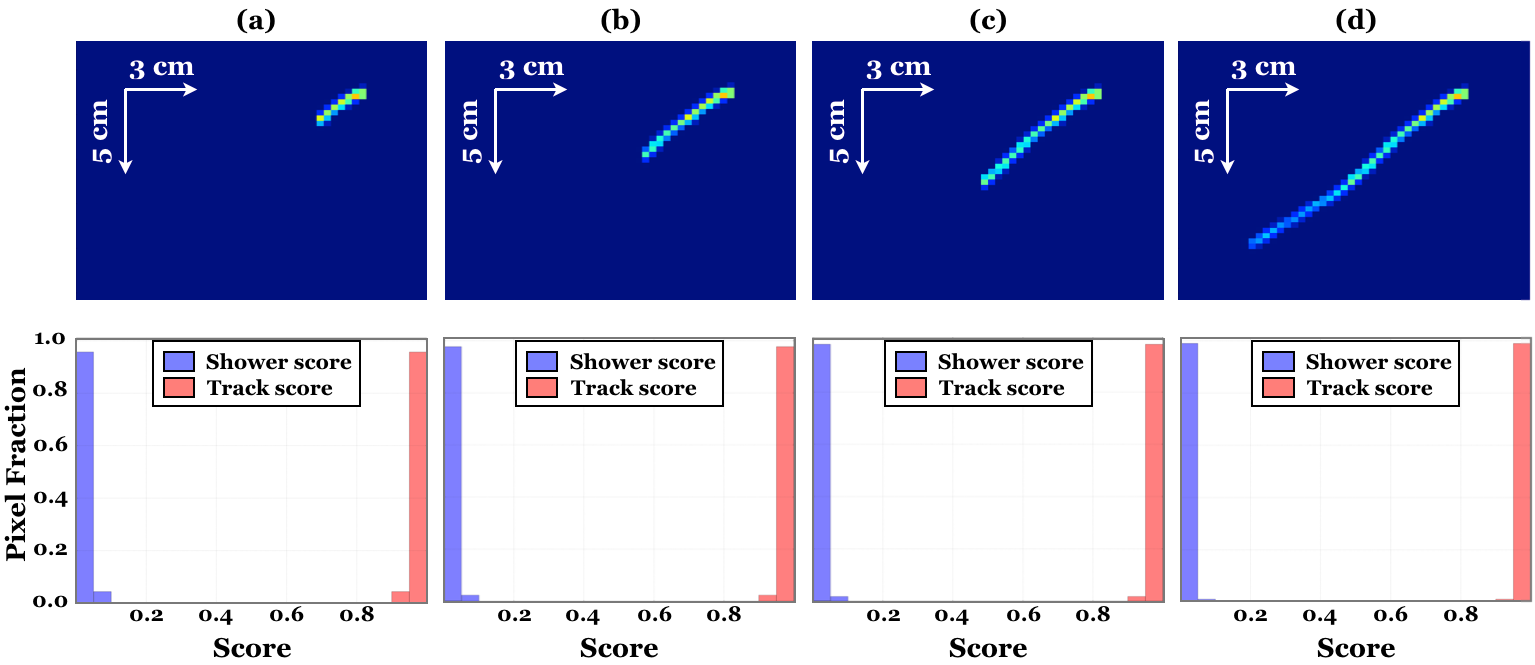}
\caption{Stopping muon track in Region~2 of Figure~\ref{fig:OneEventExample_Michel_01} where all pixels in the image are masked except for  the small portion of the muon track next to its stopping point shown in the images above. The lower row shows normalized track and shower score distributions for all non-zero pixels in the image.}
\label{fig:Region2_Michel}
\end{figure*}
\subsubsection{Bragg Peak}
A stopping muon increases its energy deposition density, $dE/dx$, as it loses momentum and near the stopping point has the highest dE/dx called the Bragg peak.  This increasing $dE/dx$ is a useful signature to identify a stopping muon~\cite{UBMichelPaper} and therefore make a decision that a trajectory is track-like.
In Figure~\ref{fig:Region1_Michel}, we show that the network struggles with a straight, minimum ionizing track-like trajectory of relatively few pixels. Figure~\ref{fig:Region2_Michel} shows Region~2 of Figure~\ref{fig:OneEventExample_Michel_01}, near the stopping muon's Bragg peak point, where we masked the rest of muon trajectory and the entire electron charge depositions. Track and shower score distributions are well separated at all track lengths. This is a distinct feature from Figure~\ref{fig:Region1_Michel}. We therefore conclude that the network is keying on an increasing $dE/dx$, or a high $dE/dx$, to classify a straight-line-like topology into track-like with high confidence even if the length of such a trajectory is down to several pixels. 

\begin{figure*}[t]
\centering
   	\includegraphics[width=0.75\textwidth]{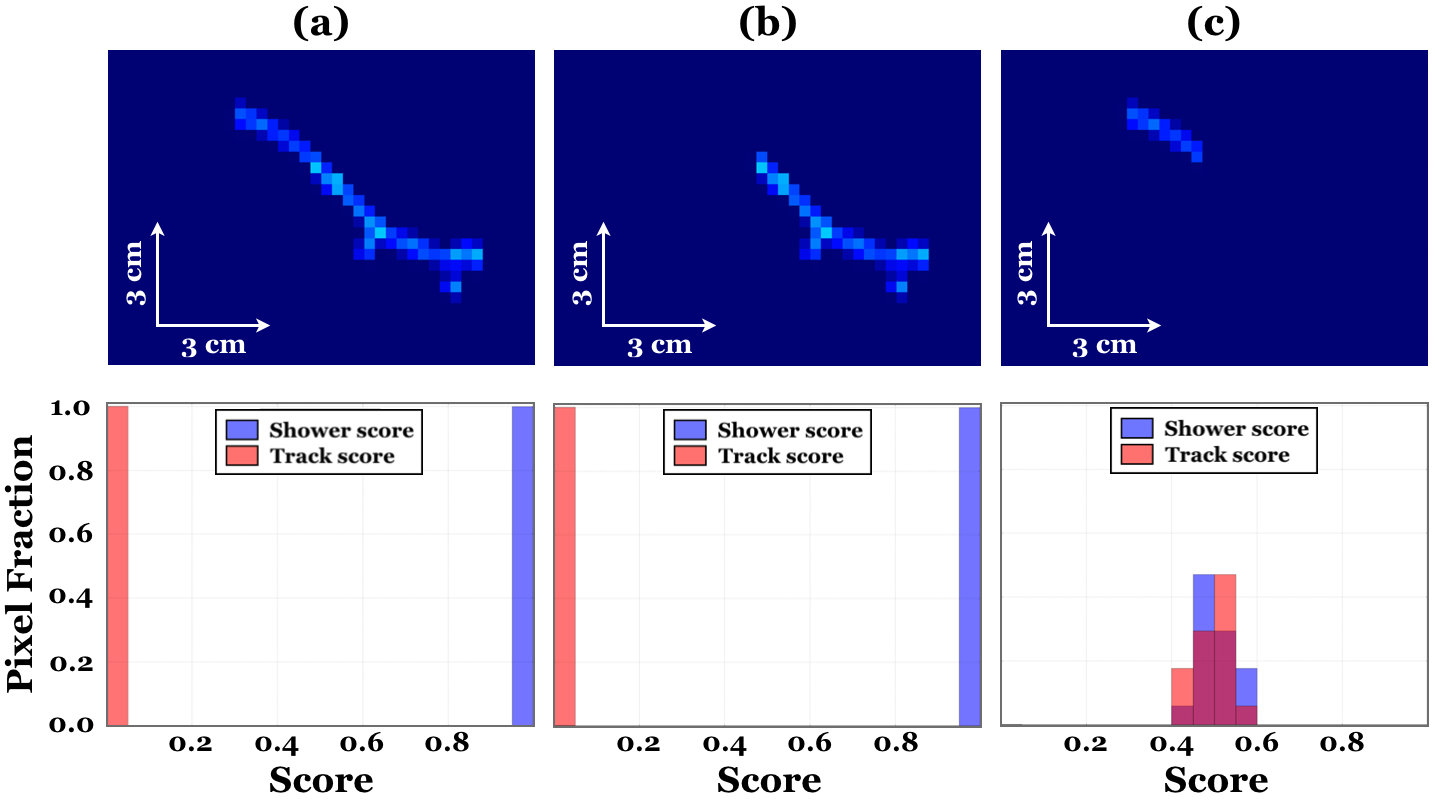}
\caption{Michel electron in Region~3 of Figure~\ref{fig:OneEventExample_Michel_01} where all pixels in the image are masked except for a portion of an electron trajectory shown in the images above. The lower row shows normalized track and shower score distributions for all non-zero pixels in the image. }
\label{fig:Region3_Michel}
\end{figure*}
\begin{figure*}[t]
\centering
   	\includegraphics[width=0.75\textwidth]{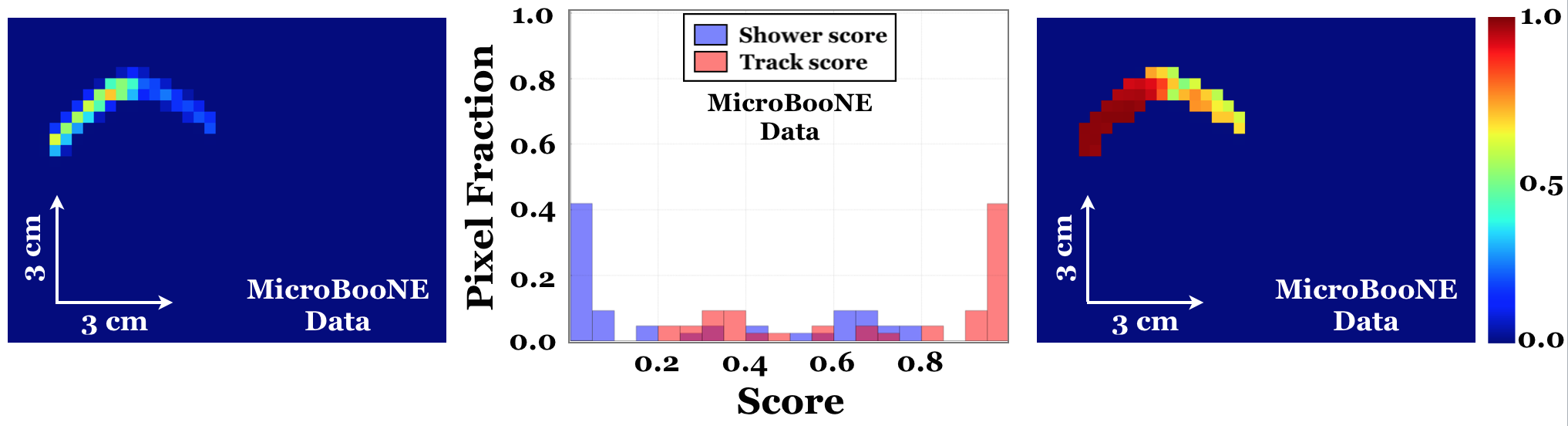}
\caption{The intersection of Region~2 and 3 of Figure~\ref{fig:OneEventExample_Michel_01} where all pixels in the image are masked except for a small portion of a Michel electron and the Bragg peak from a stopping muon. The middle histogram shows a normalized track and shower score distribution for all non-zero pixels in the image. The right figure shows the normalized score value of the classified pixel category.}
\label{fig:Region3_Brag}
\end{figure*}
\subsubsection{Low Energy Electron Shower}
One key feature of an electromagnetic shower is its non-straight line trajectory.
To test this hypothesis, we take a closer look at the Michel electron charge deposition in Region~3 of Figure~\ref{fig:OneEventExample_Michel_01}. In this event a Michel electron traveled straight for several pixels. Then it started to scatter off of other electrons before the end of the trajectory. Because of its mass, the Michel electron is minimum ionizing for most of its trajectory. We investigate how the network's confidence varies if we separate the initial straight, minimum ionizing trajectory of a Michel electron from the remaining image.

Figure~\ref{fig:Region3_Michel} shows that, where the Michel electron picture is complete, all pixels are identified strongly as a shower. 
We then mask the first several pixels that look like a track of a minimum ionizing muon (Figure~\ref{fig:Region3_Michel}(b)). The network's confidence remains very strong in this region. 
We also show the network's response to the first several pixels of a Michel electron (Figure~\ref{fig:Region3_Michel}(c)). The network is entirely uncertain whether this is a track or a shower in this case.

Finally, we investigate the intersection of Region 2 and 3 by adding the final pixels of the Bragg peak of the stopping muon to the first several pixels of the Michel electron, as shown in Figure~\ref{fig:Region3_Brag}
We add a {\it heat map} which shows the score for non-zero pixels for the classified category. In the heat map we observe the dark-red region, corresponding to the score value 1, and the other region in yellow/orange color which corresponds to weaker classification scores. We also show the track and shower score distribution of all non-zero pixels. The network's confidence level remains high on classifying the Bragg peak as track-like. 
Secondly, by comparing the score distribution of Figure~\ref{fig:Region3_Brag} and the score distribution in Figure~\ref{fig:Region3_Michel}(c), we conclude that the beginning portion of the Michel electron now has a higher likelihood to be classified as shower-like. This suggests that the network is using the presence of the Bragg peak in the image to improve the classification of the minimum ionizing straight trajectory that starts from the end of the Bragg peak, which is otherwise ambiguous as track-like or shower-like. 


\begin{figure}[t]
	\centering
    \includegraphics[width=0.45\textwidth]{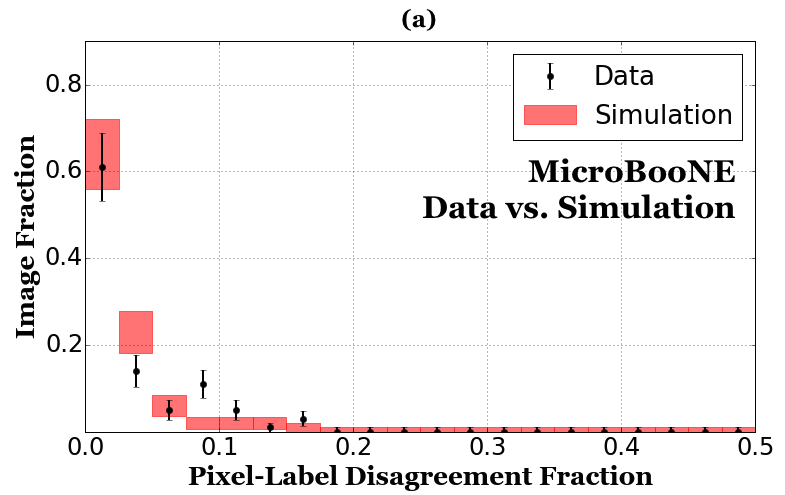}\\
    \vspace{0.1in}
	\includegraphics[width=0.45\textwidth]{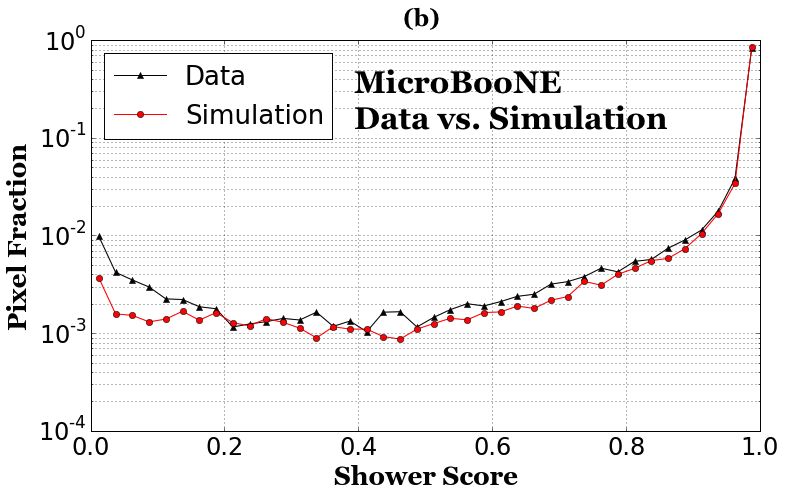}\\
    \vspace{0.1in}
    \includegraphics[width=0.45\textwidth]{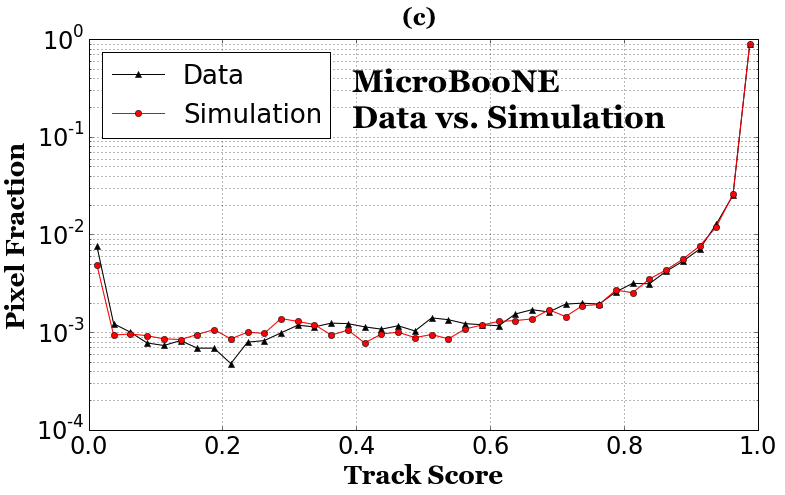}
    \caption{(a) The binned distribution of ICPF where the pixel-level labels are produced by a physicist. The Data (black) and simulation (red) distributions are area-normalized, produced from 100 CC$\pi^0$ events.  There is no event outside the shown range on the horizontal axis. (b) The normalized, binned softmax probability distributions for shower pixels by the network on data and simulation. (c) The same as (b) for track pixels.}
    \label{fig:DataMC_Score_CCPi0}
\end{figure}
\begin{figure}[t]
	\centering
    \vspace{0.1in}
	\includegraphics[width=0.45\textwidth]{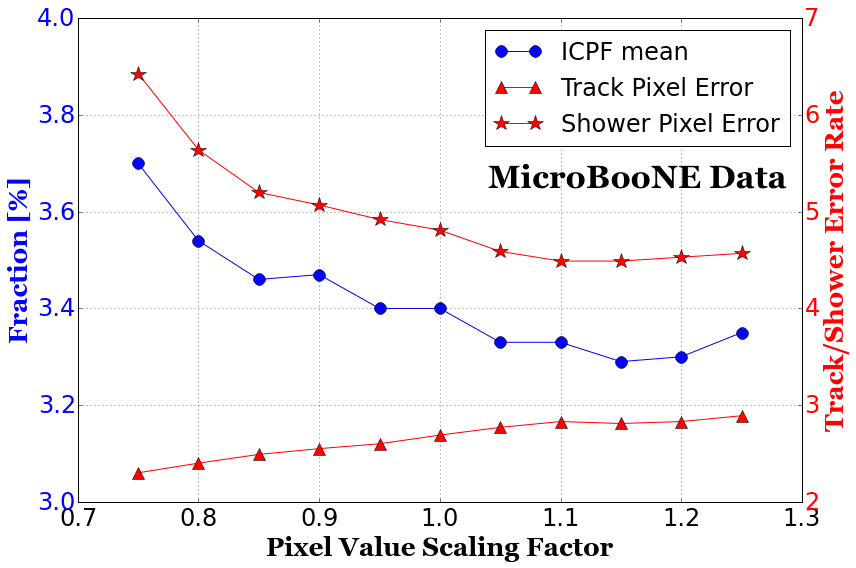}
	\caption{The ICPF mean in percent for CC$\pi^0$ data (blue) for varying pixel scaling factor shown on the horizontal axis. A category-wise physicist-network disagreement pixel fraction in percent is shown in red for track and shower pixels separately.}
	\label{fig:CCPi0ScalingResponse}
\end{figure}
\subsection{Data/Simulation Comparison Using The CC$\pi^0$ Sample}
The results of running U-ResNet on the CC$\pi^0$ image samples are summarized in Table~\ref{tab:TestAccuracyCCPi0}.
We find a similar trend as observed in the Michel electron sample, but with slightly higher disagreement rates. This is expected given that CC$\pi^0$ samples are complex images because of the higher number of particles and interactions involved. The top plot in Figure~\ref{fig:DataMC_Score_CCPi0} shows the ICPF distribution. As in the case of the Michel electron sample, data and simulation are in agreement within statistical fluctuations.  In Figure~\ref{fig:DataMC_Score_CCPi0}(b,c), we show the score distributions for the track and shower pixels labeled by a physicist. A similar trend is observed between data and simulation in both pixel categories. Four example images are shown in the Appendix with the U-ResNet output. The displayed events are visually selected by a physicist because of their particularly busy vertex activities.
\begin{table}[t]
\begin{center}
\caption{Values of the network performance metrics for the CC$\pi^0$ sample. The top row indicates the type of a sample used (simulation or data), the second shows the source of a label used for analysis, and the third shows the source of a pixel prediction. The forth and the fifth rows indicate the ICPF mean value over all samples and 90\% quantile, respectively. The bottom two rows show the mean of ICPF for track and shower pixels, separately.  Values are given as percentages.}
\vspace{0.1in}
\begin{tabular}{l||c|c|c|c}
\hline
\hline
Sample    & Data     & Simulation   & Simulation & Simulation\\
\hline
Label     & Physicist    & Physicist        & Simulation & Simulation\\ 
\hline
Prediction& U-ResNet & U-ResNet     & U-ResNet   & Physicist\\
\hline
ICPF mean & 3.4 & 2.5 & 1.8 & 2.0 \\
\hline
ICPF 90\% & 9.0 & 5.7 & 4.6 & 4.8 \\
\hline
Shower    & 4.8 & 3.4 & 3.0 & 2.6 \\
\hline
Track     & 2.7 & 2.4 & 2.2 & 2.9 \\
\hline
\hline
\end{tabular}
\label{tab:TestAccuracyCCPi0}
\end{center}
\end{table}

Following the analysis of the Michel sample, we investigate how the scaling of pixel values in the data image affects the network performance. The result is shown in Figure~\ref{fig:CCPi0ScalingResponse}. We find that the ICPF values have small variation among the scaling factors applied in the study. 
This suggests that, although the effect is small, at the 1\% level for a 25\% pixel value scaling factor, mismatched signal strength in data does affect the network's response.  


\begin{figure*}[p]
\centering
   	\includegraphics[width=1.0\textwidth]{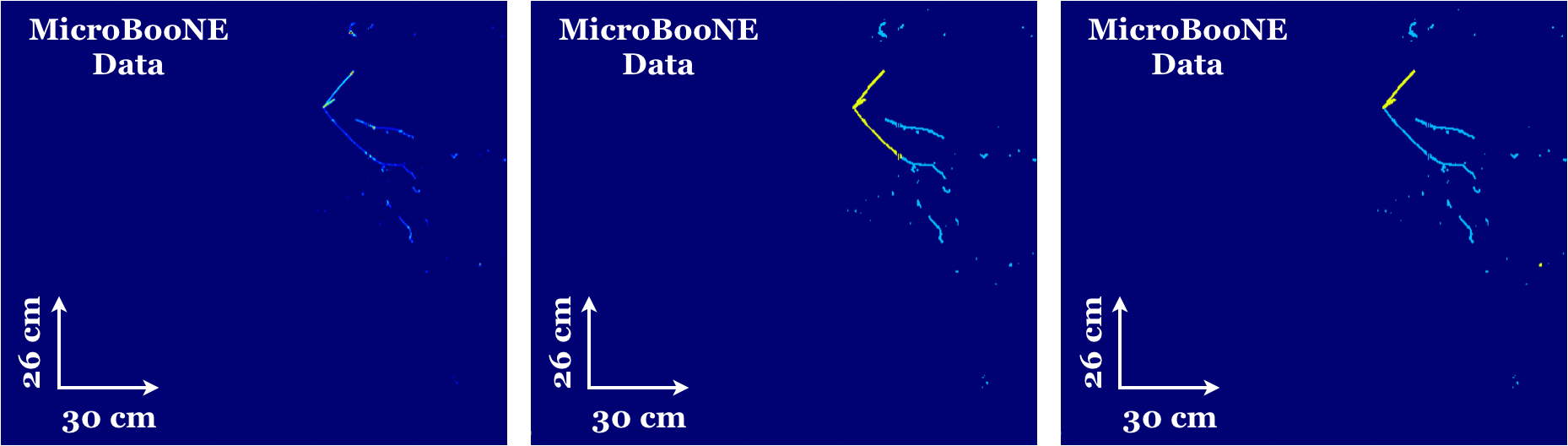}\\
   	\includegraphics[width=1.0\textwidth]{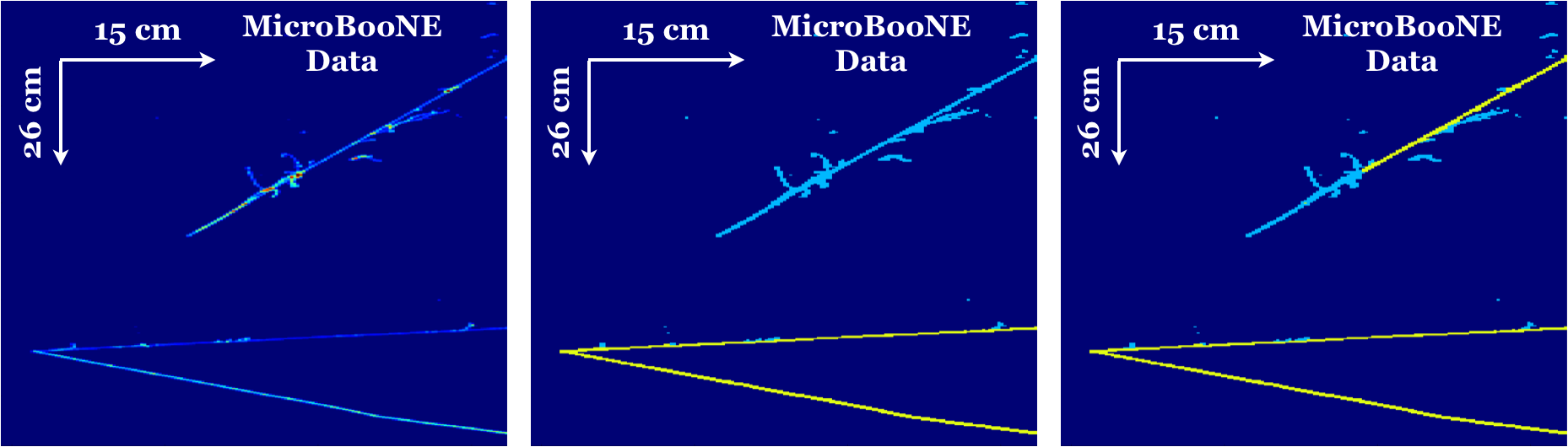}\\
   	\includegraphics[width=1.0\textwidth]{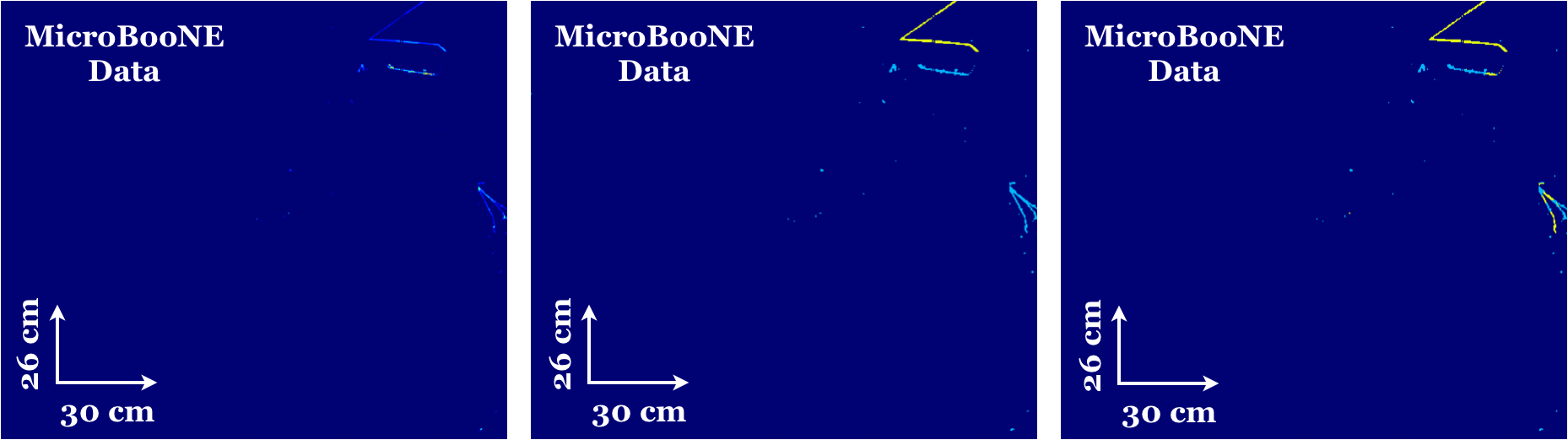}\\
   	\includegraphics[width=1.0\textwidth]{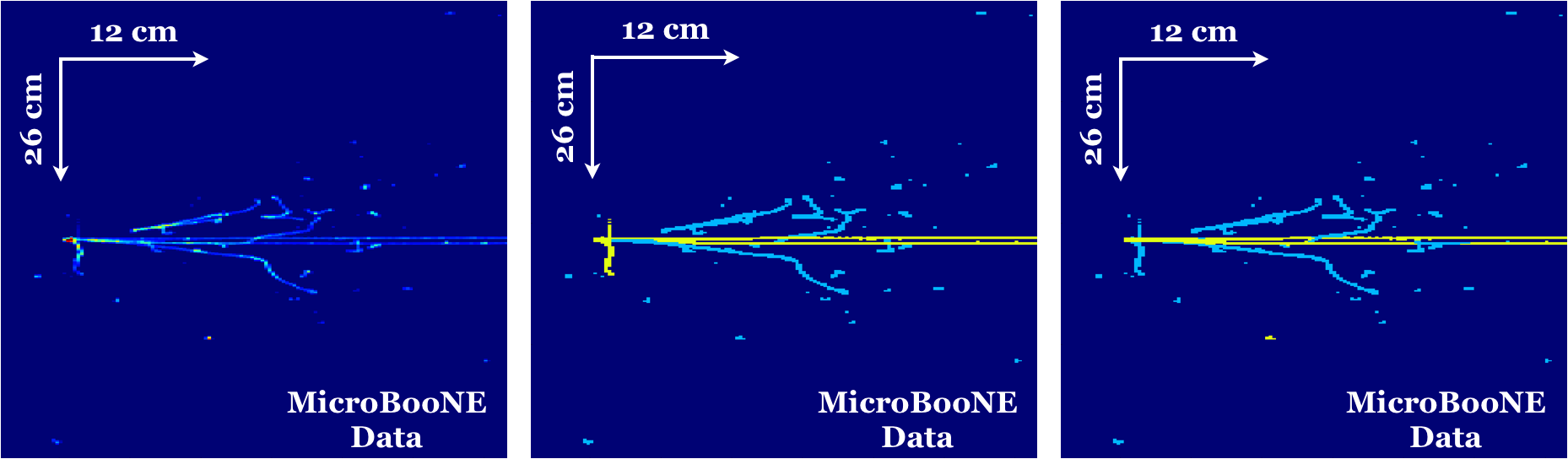}
\caption{Four example CC$\pi^0$ events with highest ICPF values using physicist generated pixel labels. Left: input images to the network. Middle: track (yellow) and shower (cyan) using physicist generated labels. Right: track (yellow) and shower (cyan) labels predicted by the network.}
\label{fig:ExampleDisplay_CCPi0Bad}
\end{figure*}
\subsection{Disagreement Between U-ResNet and Physicist Labeling for the CC$\pi^0$ Sample}
The CC$\pi^0$ events present far richer topologies than the Michel events, and we do not attempt to perform the pixel-masking and the deduction exercise to learn how the network works in this sample. Instead, we study the CC$\pi^0$ data events where the disagreement between physicist and network labeling is largest. Four such events are identified and shown in Figure~\ref{fig:ExampleDisplay_CCPi0Bad}. The four events shown are ordered by level of disagreement rates of 0.166, 0.166, 0.162, and 0.125, respectively.  In the example shown on the top of the figure, the disagreement is mainly in a long track-like trajectory originating from the interaction vertex. While a physicist analyzer decided this is a track, it could also be a minimum ionizing electron that should be classified as a shower. The second display from the top shows the network's attempt to separate a track-like trajectory that is present inside a high energy shower. 
In the third image, a large portion of particle trajectory is invisible due to unresponsive region of the detector running vertically toward the right of this image. This makes it difficult to analyze the remaining particle trajectories where the U-ResNet mixes track and shower pixel decisions for the same trajectory.
Finally, in the bottom image, the network predicts two track-like trajectories coming from the interaction vertex while the physicist analyzer merged two tracks into one near the vertex. The region around the interaction vertex is complicated by energy deposition from a shower, which makes it difficult to determine which decision is correct. 
The performance of the network for these maximal disagreement events supplements our understanding from the qualitative Michel data analysis and documents some of the ways that the network fails in categorizing pixels.

\section{Conclusion}
In this paper we have presented the first application of a deep semantic segmentation network, U-ResNet, to perform track/shower separation at the pixel-level for LArTPC images. We explore training techniques including transfer learning and pixel-wise error weighting methods. Our software tools and algorithms to store and apply the pixel-level labeling are made available in  Refs~\cite{LArCV,uboonecode06260113}. 

U-ResNet achieved an average ICPF of 6.0\% and 3.9\% benchmarked with 20,000 images of $\nu_e$ and $\nu_\mu$ interactions, respectively,  simulated with realistic neutrino beam information. The same network achieved an average ICPF of 3.9\% for 1e1p-LE events in which electrons have a uniform momentum distribution from 30 to 100~MeV/c, and protons from 300 to 450~MeV/c. The average ICPF is found to be 2.3\% for 1$\mu$1p-LE events which include protons with the same uniform momentum distribution and momentum range, and muons in a momentum range of 85 to 175~MeV/c. 


We quantified and validated U-ResNet, trained purely on simulated image samples, on LArTPC images from real detector data. We calculated the fraction of incorrect pixel labeling between U-ResNet and a physicist analyzer and found an average disagreement fraction of 1.9\% and 3.4\% for Michel electron data and CC$\pi^0$ data, respectively. The same analysis was performed using simulation samples, and we found that the level of disagreement is consistent for data and simulation samples.  This is the first time such validations have been shown on real LArTPC data. From a qualitative analysis on the Michel electron data we conclude that the network is focusing on intuitively reasonable physics features in the image. The successful application of a semantic segmentation network on LArTPC data is an important milestone toward developing a full LArTPC data reconstruction chain using a deep neural network.

\section{Acknowledgement}
This document was prepared by the MicroBooNE collaboration using the resources of the Fermi National Accelerator Laboratory (Fermilab), a U.S. Department of Energy, Office of Science, HEP User Facility. Fermilab is managed by Fermi Research Alliance, LLC (FRA), acting under Contract No. DE-AC02-07CH11359.  MicroBooNE is supported by the following: the U.S. Department of Energy, Office of Science, Offices of High Energy Physics and Nuclear Physics; the U.S. National Science Foundation; the Swiss National Science Foundation; the Science and Technology Facilities Council of the United Kingdom; and The Royal Society (United Kingdom).  Additional support for the laser calibration system and cosmic ray tagger was provided by the Albert Einstein Center for Fundamental Physics, Bern, Switzerland.

%
%

\section*{Appendix}
\label{sec:appendix:display}
\begin{figure*}[p]
\centering
	\includegraphics[width=1.0\textwidth]{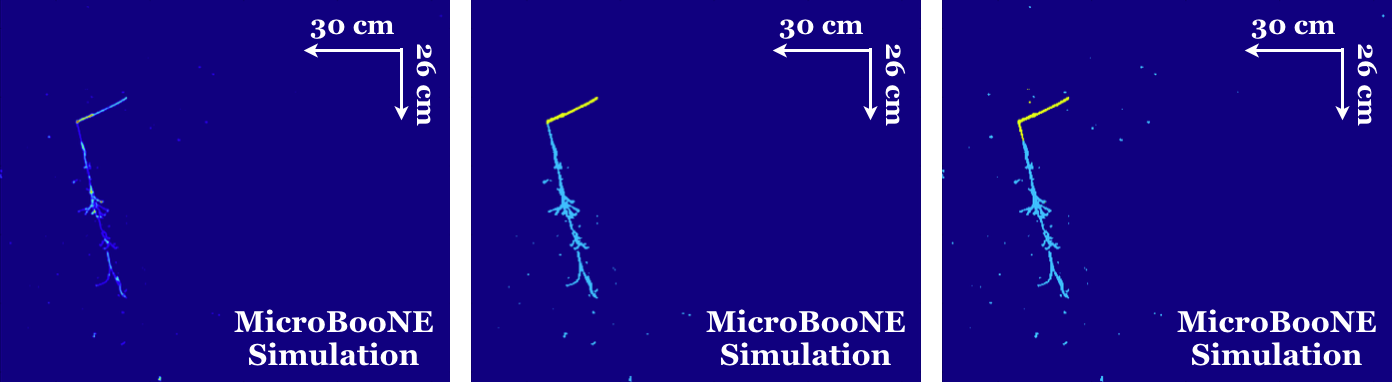}\\
	\includegraphics[width=1.0\textwidth]{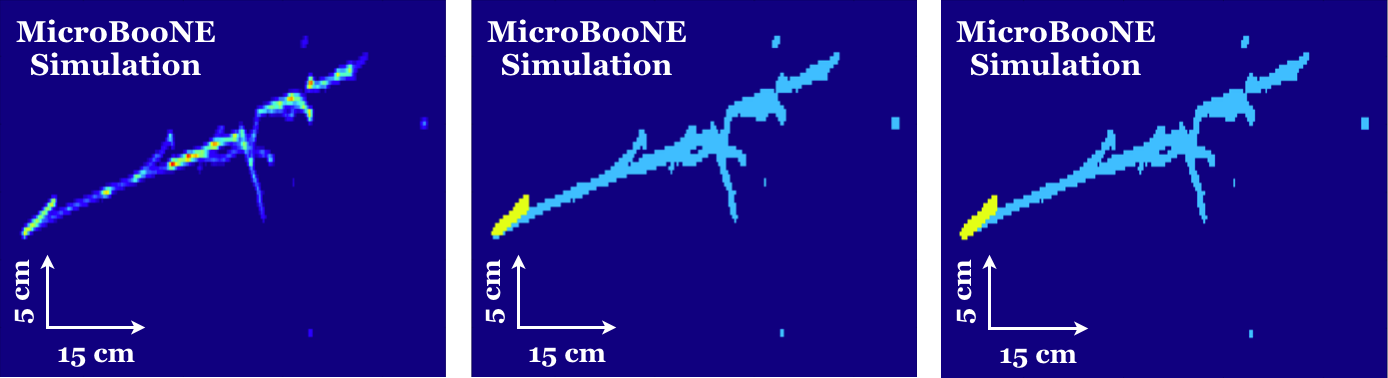}\\
  	\includegraphics[width=1.0\textwidth]{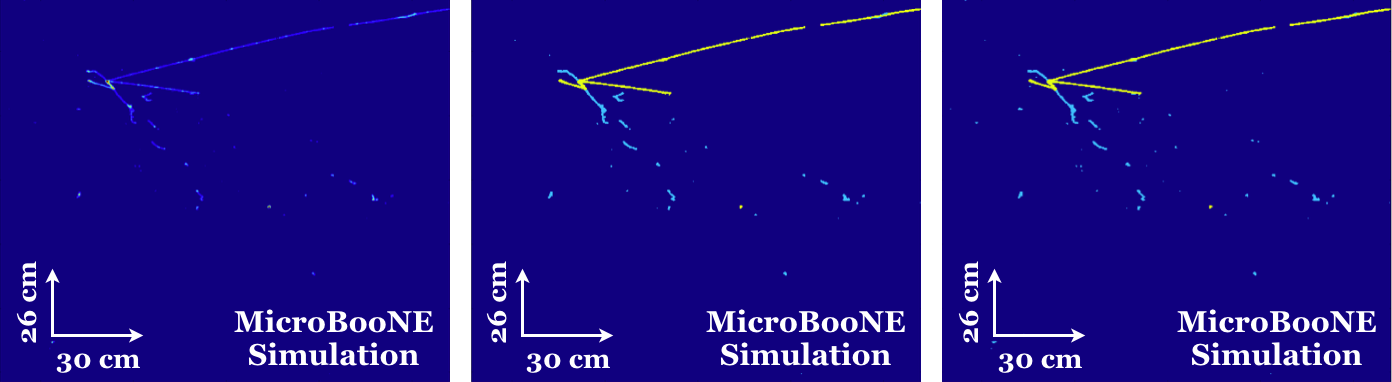}\\
	\includegraphics[width=1.0\textwidth]{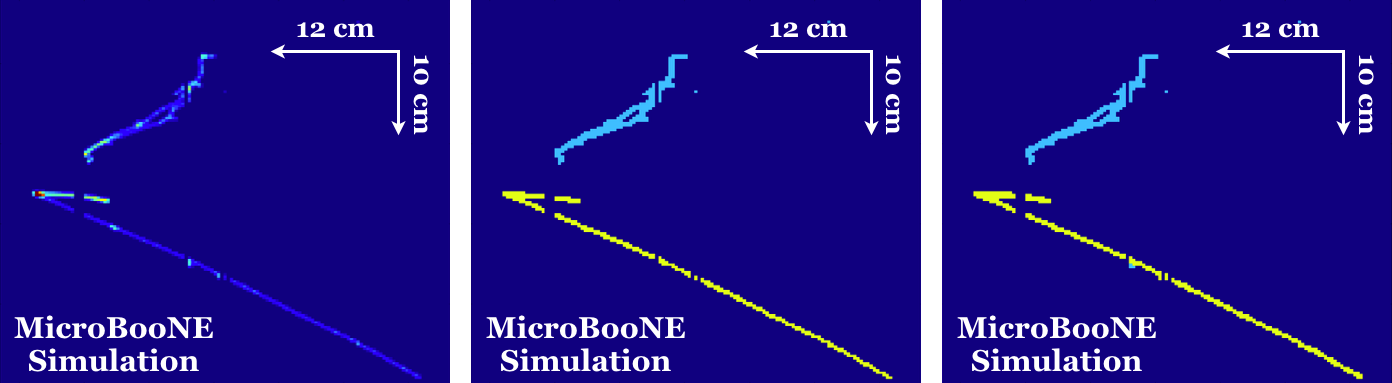}\\
\caption{Event displays of $\nu_e$ (upper two rows) and $\nu_\mu$ interactions (lower two rows). The left column images are inputs to the network. The middle column shows labeled images based on simulation information. Track pixels are masked in yellow and shower pixels are in cyan. The right column shows the output of the network.}
\label{fig:ExampleMCInference}
\end{figure*}

\begin{figure*}[p]
\centering
   	\includegraphics[width=\textwidth]{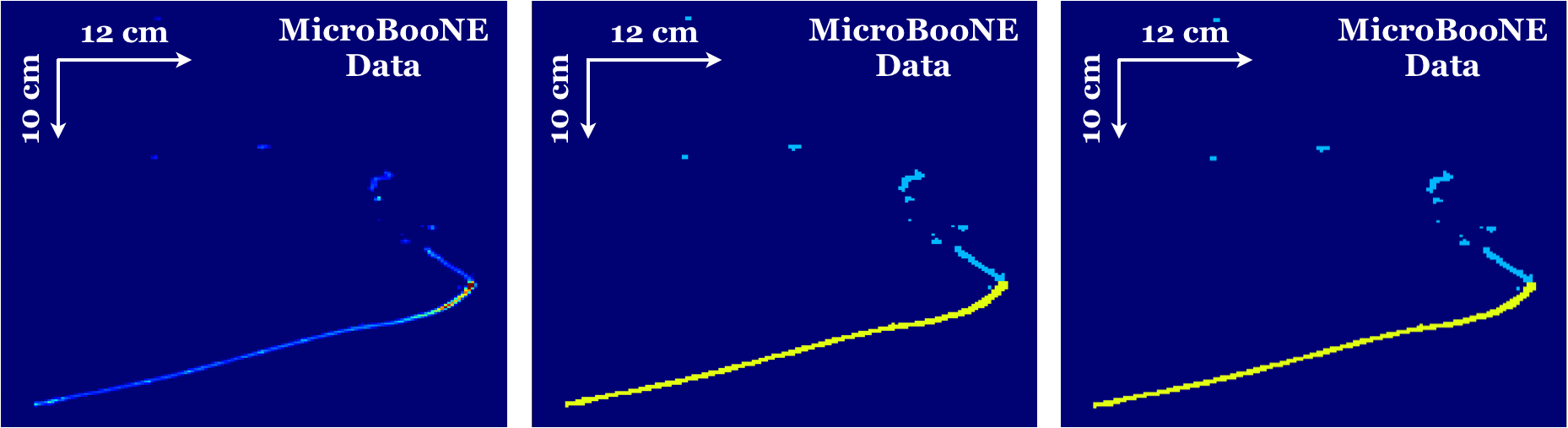}\\
   	\includegraphics[width=\textwidth]{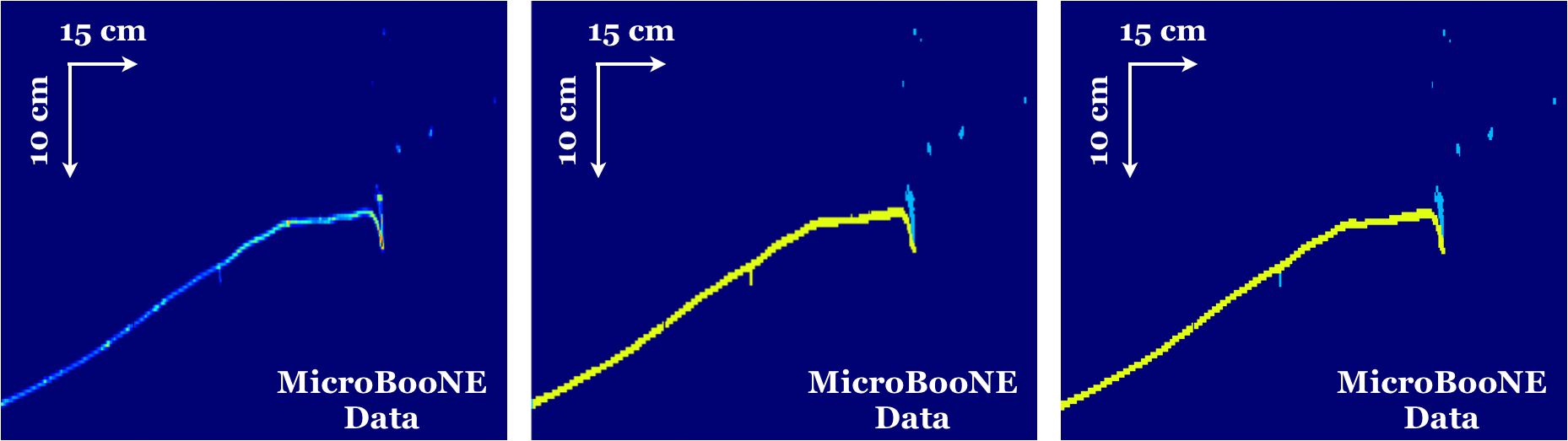}\\
   	\includegraphics[width=\textwidth]{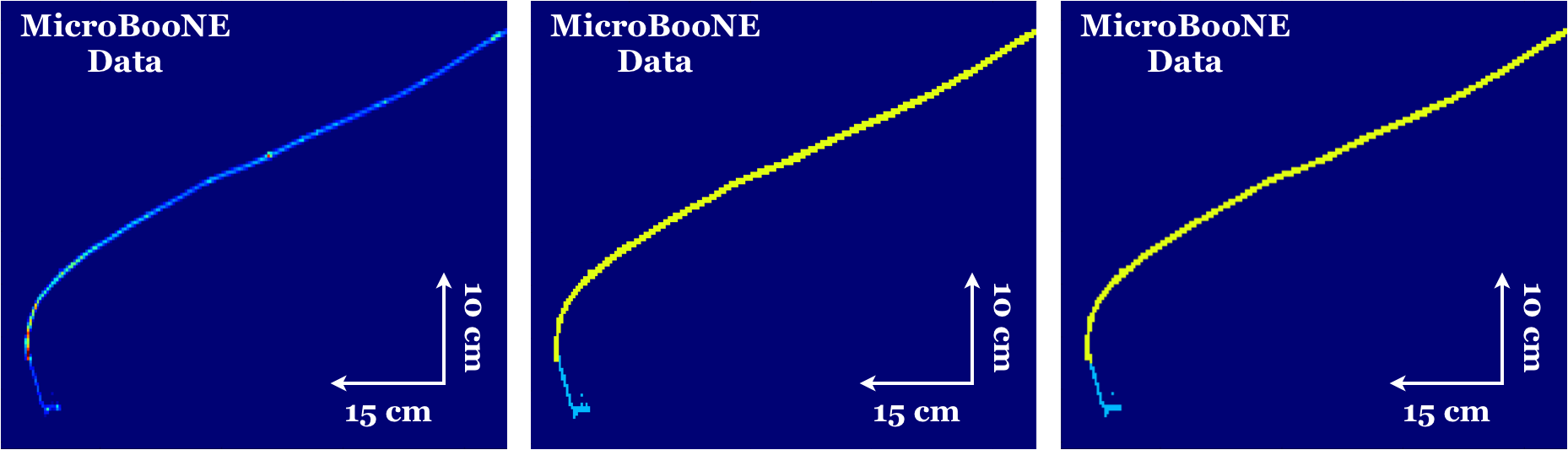}\\
   	\includegraphics[width=\textwidth]{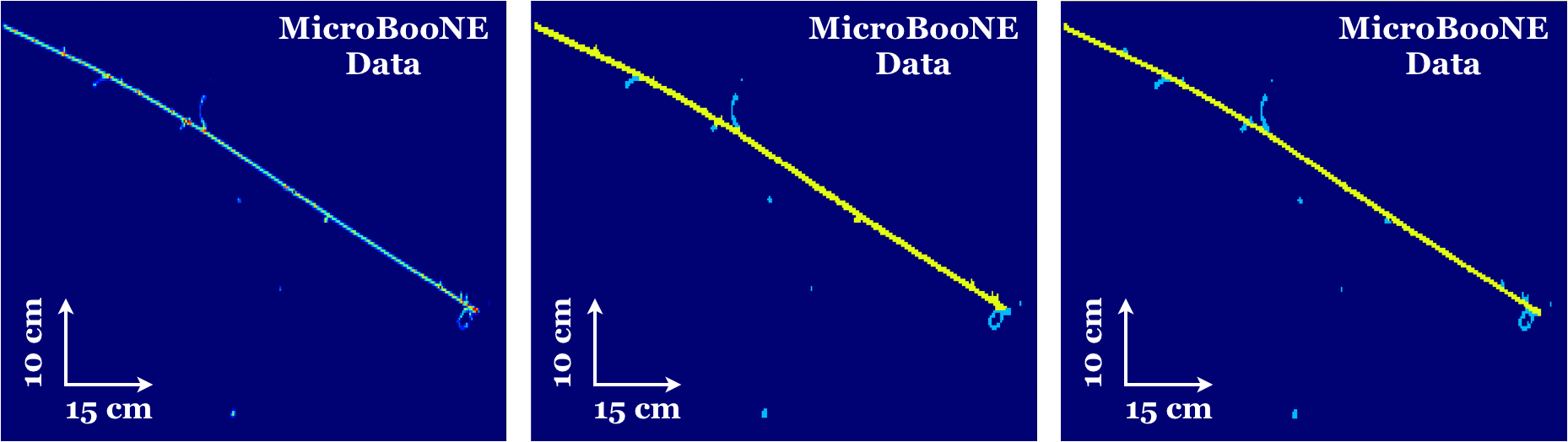}
\caption{Michel electron event displays from real detector data. Left: input images to the U-ResNet. Middle: track (yellow) and shower (cyan) physicist labels. Right: track (yellow) and shower (cyan) labels predicted by the network.}
\label{fig:ExampleDisplay_Michel}
\end{figure*}

\begin{figure*}[p]
\centering
   	\includegraphics[width=1.0\textwidth]{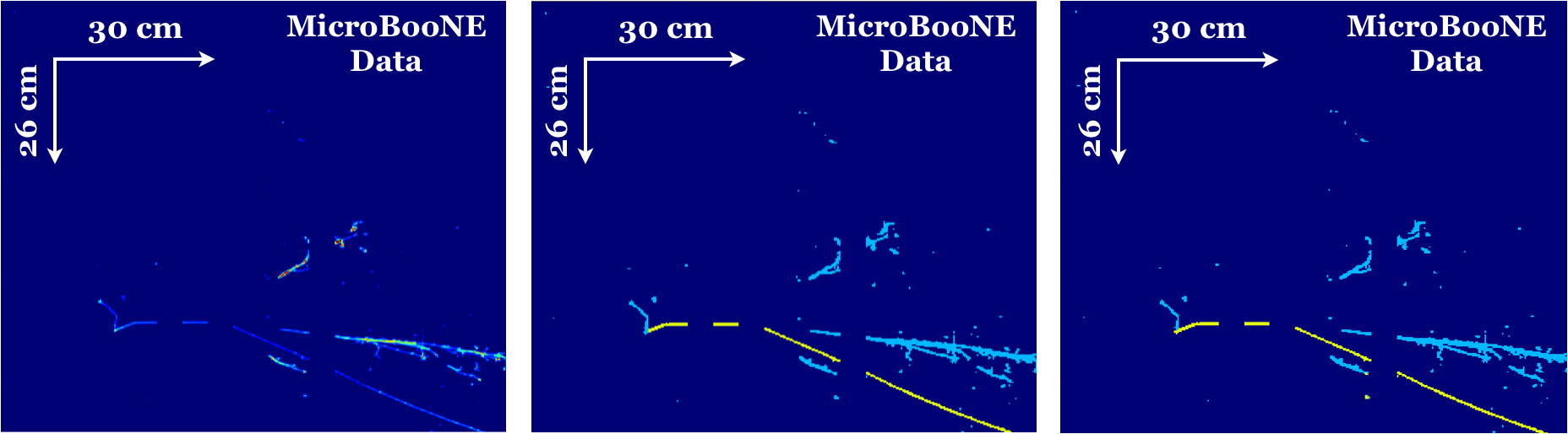}\\
   	\includegraphics[width=1.0\textwidth]{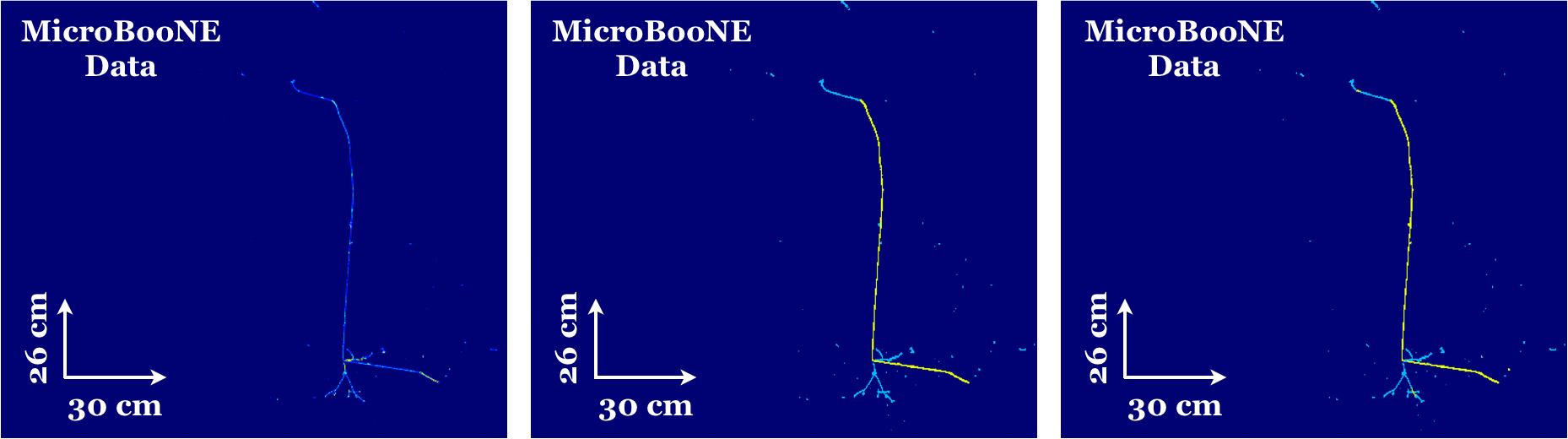}\\
   	\includegraphics[width=1.0\textwidth]{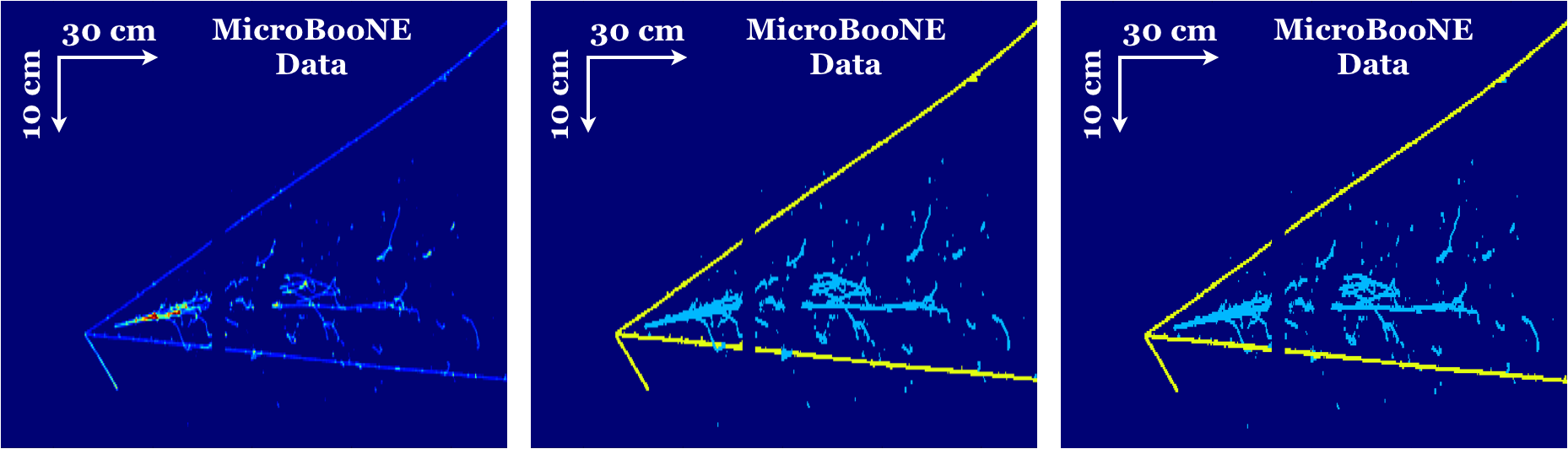}\\
   	\includegraphics[width=1.0\textwidth]{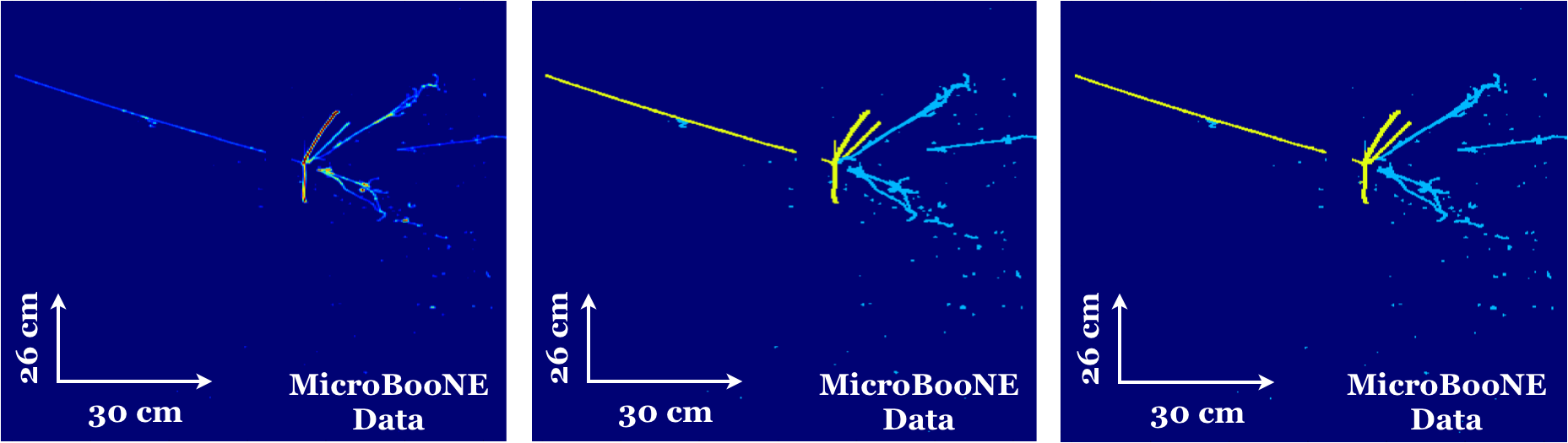}
\caption{Neutrino event displays from CC$\pi^0$ candidate detector data selected based on activity around the interaction vertex. Left: input images to the network. Middle: track (yellow) and shower (cyan) physicist labels. Right: track (yellow) and shower (cyan) labels predicted by the network.}
\label{fig:ExampleDisplay_CCPi0Busy}
\end{figure*}

In this appendix we show example event displays of simulation and data. Figure~\ref{fig:ExampleMCInference} shows simulated $\nu_e$ and $\nu_\mu$ interactions. Figure~\ref{fig:ExampleDisplay_Michel} shows a stopping muon with a decay Michel electron. Figure~\ref{fig:ExampleDisplay_CCPi0Busy} shows ``busy'' CC$\pi^0$ candidate events visually selected including those with a particle multiplicity greater than 4. In all of these event displays, gaps in tracks and showers are due to unresponsive wires. Overall, we observe good agreement between the simulation information and output from the network track-shower pixel labeling in diverse event types.

%
%

%
%

\end{document}